\documentclass[10pt,a4paper]{article}

\usepackage[latin1]{inputenc}
\usepackage{amsmath}
\usepackage{amsfonts}
\usepackage{amssymb}
\usepackage{graphicx}
\usepackage{hyperref}
\usepackage{color}
\usepackage{multicol}
\usepackage{lipsum}
\usepackage{mwe}
\title{On the Difference between Physics and Biology: \\Logical Branching and Biomolecules}

\author{George F R Ellis\\
Mathematics Department, University of Cape Town\\
and\\
Jonathan Kopel
\\ Texas Tech University Health Sciences Center (TTUHSC)}
\begin{document}
\maketitle
\begin{abstract}
Physical emergence - crystals, rocks, sandpiles, turbulent eddies, planets, stars - is fundamentally different from biological emergence - amoeba, cells, mice, humans - even though the latter is based in the former. This paper\footnote{This paper is based in an FQXI \href{http://fqxi.org/community/forum/topic/2769}{essay} written by one of us (GE), and is a companion to a paper with Philippe Binder  on the relation between computation and physical laws \cite{BinEll18}.} points out that an  essential difference is that as well as involving physical causation, causation in biological systems  has a logical nature at each level of the hierarchy of emergence, from the biomolecular level up.   
The key link between physics and life enabling this to happen is provided by biomolecules, such as voltage gated ion channels, which enable branching logic to emerge from the underlying physics and hence enable logically based cell processes to take place in general,  and in neurons in particular. These molecules can only have come into being via the contextually dependent processes of natural selection, which selects them for their biological function. A further major difference is between life in general and intelligent life. We characterise intelligent organisms as being engaged in deductive causation, which enables them to transcend the physical limitations of their bodies through the power of abstract thought, prediction, and planning. Ultimately this is enabled by the biomolecules that underlie the propagation of action potentials in neuronal axons in the brain.
\end{abstract}
%\begin{abstract}FQXI Essay for  \href{http://fqxi.org/community/forum/category/31425}{Wandering Towards a Goal} Essay contest.
%\end{abstract} 
\section{Physics vs Biology}\label{sec:function}

	How	can a	universe	that is ruled	by	natural	laws	give rise to	function and purpose? This issue arises in the context of the ongoing debate about emergence and reductionism in science \cite{Emerge1,Emerge2,Ell16} and attains current relevance because of a series of recent %popular books in effect proposing that biology has no special significance as it is entailed by physics, these ideas being buttressed by some technical
	 papers deriving aspects of biological functioning arising from purely physical considerations 
	 \cite{Cockell,England1,England2}. These aspects  are 
	 certainly of interest and importance, but they do not get 
	 at the essence of biological existence. Our viewpoint is 
	 that physics and biology are essentially different, even 
	 though physics underlies biology. This agrees with the 
	 viewpoint of Ernst Mayr, who talks of dual  causation in 
	 biology (\cite{Mayr2004}:89-90), where purely physical 
	 causation occurs at the lower levels but essentially 
	 different biological causation due to developmental 
	 programs also occurs, leading to teleonomic behaviour 
	 (\cite{Mayr2004}:51-57). We will generalise that argument 
	 to a broader characterisation of the physics-biology 
	 difference as being due to the fact that biological 
	 causation is based in logical branching shaped by 
	 context, enabled in physical terms by the nature of 
	 particular proteins. This contextual emergence 
	 	 \cite{ContextEmerge} is a form of strong emergence. This proposal agrees with the view of Davies 
	 and Walker \cite{DavWal16}, who see information as a causal 
	 agent in biology. The use of information in a contextual 
	 way that they envisage, for example in their analysis of 
	 the Boolean network model of the gene regulatory network
	 of the fission yeast \textit{S. Pombe} cell cycle, relies 
	 on the kind of branching logic discussed in this paper.
	 \\
	
	 The basic point is that the key difference between physics and biology is existence of function or purpose in biology. 	Whether or not a human observer exists, the natural laws continue to operate as they do in a purposeless way with no function implied. There is, in the standard scientific interpretation, no purpose in the existence of the Moon\footnote{It is true that the existence of the Moon was probably essential for the origin of life as we know it, so one might claim that the purpose of the Moon was to enable life on Earth to emerge from the sea to dry land through creating tides. However the Moon is unaware of this effect: it was, as far as the Moon was concerned, an unintended by product of its orbital motion round the Earth.} or an electron or in a collision of two gas particles. 
By contrast, there is purpose or function in all life, as pointed out by Hartwell \textit{et al} \cite{Hart99}:
\begin{quote}
``\textit{Although living systems obey the laws of
physics and chemistry, the notion of
function or purpose differentiates biology
from other natural sciences. Organisms
exist to reproduce, whereas, outside religious
belief, rocks and stars have no purpose.
Selection for function has produced the living
cell, with a unique set of properties that
distinguish it from inanimate systems of
interacting molecules. Cells exist far from
thermal equilibrium by harvesting energy
from their environment. They are composed
of thousands of different types of molecule.
They contain information for their survival
and reproduction, in the form of their DNA}''.
\end{quote}
Functional talk is a 
contested area in the philosophy of biology \cite{Millikan,Neander,Godfer-Smith1994,Amundson}. Here, we will just make two 
remarks that should be an adequate foundation 
for what follows. Firstly, one cannot sensibly talk about physiology of living systems without talking about 
function or purpose: the heart exists in 
order to circulate blood (\cite{AnimalPhysiology}:476-510), pacemaking cells 
exist in order to determine the rhythm of the 
heart, blood exists in order to transport 
oxygen, a key function of mitochondria in eukaryotes is to provide energy to the cell by converting ingested sugars into ATP (\cite{AnimalPhysiology}:74), and so on \cite{RhoPfl89}. Thus in \textit{Animal Physiology} \cite{AnimalPhysiology}, Randall \textit{et al} write
\begin{quote}
``\textit{Animal physiology focuses on the functions of tissues, organs, and organ systems in multicellular animals... Function flows from structure... A strong relationship between structure and function occurs at all levels of biological organisation.''} (\cite{AnimalPhysiology}:3,6)
\end{quote}
 This is taken for granted by working biologists, as in the Hartwell \textit{et al}  quote above.
Secondly, as 
indicated in the above quote, functional talk 
in biology is vindicated by Darwinism, as 
discussed \textit{inter alia} by Ernst Mayr  
\cite{Mayr2002,Mayr2004} and Kampourakis 
\cite{Kampourakis}. 
According to 
Kampourakis, ``\textit{function} is the role 
of a component in the organization of a 
system. The functions of the parts and 
activities of organisms in enabling their 
continued existence are ``biological 
functions'' or ``biological roles'' (the 
function of the wing of an eagle is to enable 
flight, but the function of the wing of a 
penguin is enabling swimming)'' 
(\cite{Kampourakis}:221) According to 
Godfrey-Smith \cite{Godfer-Smith1994}, 
supported by Birch (\cite{Bir2017}:22),  
``Biological functions are dispositions or 
effects a trait has which explain the recent
maintenance of the trait under natural 
selection''.  According to Randall \textit{et al} \cite{AnimalPhysiology},
 \begin{quote}
 \textit{``The physiology of an animal is usually 
 very well adapted to the environment that the 
 animal occupies, thereby contributing to its 
 survival. Evolution by natural selection is the 
 accepted  explanation for this condition, called 
 adaptation. .. A physiological process is adaptive 
 if it is present at high frequency in the 
 population because it results in a higher 
 probability of survival and reproduction than 
 alternative processes''} (\cite{AnimalPhysiology}:7).
 \end{quote}  Thus function $\alpha$ of a 
trait $A$, or of a physiological system $S$ 
that enables $A$,  exists in individuals in a 
population because having property %$X$
 $\alpha$ tends to increase differential  
 reproduction and survival rates, and so 
 natural selection \cite{Darwin} 
 is a mechanism for developing property 
 $\alpha$ over recent evolutionary timescales 
 \cite{Godfer-Smith1994,Mayr2002,Mayr2004,Bir2017}, resulting in genotypes $X$ that will 
 lead to existence of property $\alpha$ in 
 individuals through contextual developmental 
 processes \cite{Kampourakis,Noble16}.  
  Additionally intention comes into play in the case 
 of conscious animals, in particular when  
 purposive behaviour (\cite{Mayr2004}:57), perhaps including  deductive causation (Section \ref{sec:6}),  
  occurs. Its  
 emergence is based on the reliable  functioning of the  underlying physiological systems \cite{AnimalPhysiology,beimGraben}. Finally, what is life? Our view will be (cf.\cite{Hart99}) that a living system is a material system that exhibits function or purpose as discussed above. \\

The key issue we address in this paper is, how does purpose or function emerge from physics on developmental and functional  timescales? How does deterministic physics lead to logical branching aimed at enabling a function? 
At the macro level, in higher animals and human beings, this occurs via adaptive neural networks \cite{Kanetal13} and physiological systems \cite{RhoPfl89}. At the micro level, through epigenetic effects in cell development \cite{GilEpe09} via   \href{https://en.wikipedia.org/wiki/Gene_regulatory_network}{gene regulatory networks} \cite{GilEpe09} and through adaptive effects in \href{https://en.wikipedia.org/wiki/Cell_signaling}{signal transduction networks} \cite{JanYaf06} and \href{http://www.mind.ilstu.edu/curriculum/neurons_intro/neurons_intro.php}{neural networks} \cite{Kanetal13}. Although many have argued that these are all based only on the lower levels through interactions of specific molecules, particularly  \href{https://en.wikipedia.org/wiki/Protein}{proteins} \cite{PetRin09} and \href{https://www2.chemistry.msu.edu/faculty/reusch/virttxtjml/nucacids.htm}{nucleic acids} \cite{Wat13}%\footnote{As many readers of this essay will be physicists rather than biologists, hyperlinks are included in the text to help clarify biological concepts introduced} ]
, in fact biological systems are as much shaped by  contextual effects  as by  bottom-up emergence \cite{Noble16}. These molecules are key, but they get their effectiveness because of the context in which they exist. As Denis Noble argued,
\begin{quote}
	\textit{``Genes, as DNA sequences, do not of course form selves in any ordinary sense. The DNA molecule on its own does absolutely nothing since it reacts biochemically only to triggering signals. It cannot even initiate its own transcription or replication. It cannot therefore be characterised as selfish in any plausible sense of the word. If we extract DNA and put it in a Petri dish with nutrients, it will do nothing. The cell from which we extracted it would, however, continue to function until it needs to make more proteins, just as red cells function for a hundred days or more without a nucleus. It would therefore be more correct to say that genes are not active causes; they are, rather, caused to give their information by and to the system that activates them. The only kind of causation that can be attributed to them is passive, much in the way a computer program reads and uses databases''} \cite{Nob11}.
\end{quote}

To be clear and concise, this paper will focus on voltage gated ion channels that underlie neuronal functioning, although the same applies for example to the active site of the enzyme  molecule which is complementary to the shape of the substrate. We will first look at the difference between the logic of physics (Section \ref{sec:2}) and the logic of biology (Section \ref{sec:3}), then at the biomolecules that make this difference possible (Section \ref{sec:4}), linking the two, and finally at the way such molecules have come into being and allowed physical processes to generate biological activity (Section \ref{sec:5}). The processes of deductive causation are discussed in Section (\ref{sec:6}). The conclusion (Section \ref{sec:7}) focuses on the plasticity and adaptation that characterise biology, and the way deductive causation distinguishes intelligent life from life in general. Appendices support the overall argument. Appendix \ref{Sec:Tech}  gives some technical notes,  Appendix \ref{sec;appendix} discusses emergent constants and parameters, Appendix \ref{sec:energy} comments on the underlying neuroenergetics, and Appendix \ref{sec:visual} on protein gradients in a squid's eye.  We take it  for granted that living systems are open non-equilibrium systems \cite{Friston}. That alone does not characterise life: famously, even a burning candle satisfies those conditions.  Something more is required.\\

Please note that we are contrasting causation in naturally occurring physical systems, such a rocks, rivers, mountains, stars, and planets, and in simple physical devices of the kind discussed by Arnold \cite{Arnold} such as rolling balls,  pendula, and tops, with causation in  biological systems. We are not considering here causation in machines, which can have much in common with biological causation \cite{Sim92}, because all machines have a purpose - indeed that is central to the concept of a machine. Thus washing machines have been designed to clean clothes, aircraft to fly, computers to do calculations, coffee machines to make coffee, and so on. The design process that leads to machines coming into being is possible because intelligent beings exist \cite{Kampourakis}, and so their existence is a consequence of the existence of life. Thus existence of these kinds of causations in designed systems should not be taken as exemplars of physical causation, but as cases of engineering causation that can have much in common with life because they are an outcome of intelligence and deductive causation (Section \ref{Sec;3_kinds}) enabled by the biological processes discussed here.

\section{Logic of Physics}\label{sec:2}
Physical laws determine evolution of a physical system in a purposeless inevitable way. Let the relevant variables be $\textbf{X}$ and the evolution dynamics, determined by energy equations (\cite{Arnold}:15-27), a force law (\cite{Arnold}:28-50), Lagrangian methods (\cite{Arnold}:55-61), or Hamiltonian mechanics  (\cite{Arnold}:65-70,165-266), be given by $H(C,\textbf{X},t)$ where the context $C$ is set by initial and boundary conditions, then that dynamical law  determines later states from earlier states \cite{Arnold}: 
\begin{equation}\label{eq:1}
\texttt{If %data 
at time $t_1$, }\, \textbf{X}= \textbf{X}(t_1),\, %\textbf{X}\, C\, 
\texttt{then at time $t_2$},\,\, \textbf{X} = H(C,\textbf{X}(t_1),t_2). 
\end{equation}
This dynamics may often be represented by suitable phase planes. Two examples are  the dynamics of classical systems such as a pendulum \cite{Arnold}, %as expressed in 
%\href{https://mathematicalgarden.wordpress.com/2009/03/29/nonlinear-pendulum/}{pendulum phase planes}
and the dynamics of celestial objects governed by gravity \cite{Binney,celestial}. %, which can also be described by phase planes.   %\href{https://www.handprint.com/ASTRO/bineye3.html}{Gravitational systems}
The key point is that in physics there are fixed interactions that cannot be altered, although we can to some extent decide what they act on and so what their outcome will be (as in the case of a pendulum or a computer). 
In the end, daily life is governed by \href{https://en.wikipedia.org/wiki/Newton's_laws_of_motion}
{Newton's laws of motion} and Galileo's \href{https://en.wikipedia.org/wiki/Equations_for_a_falling_body}{equations for a falling body}, together with \href{https://en.wikipedia.org/wiki/Maxwell's_equations}{Maxwell's equations}:\footnote{We are avoiding discussing the extra complications introduced by quantum mechanics at this point. This will be important below (Section \ref{sec:exist}).}
\begin{eqnarray}\label{eq:maxwell}
\nabla \cdot \mathbf {E} =4\pi \rho, \,\,
&\nabla \times \mathbf {E} =-{\frac {1}{c}}{\frac {\partial \mathbf {B} }{\partial t}},\\
 \nabla \cdot \mathbf {B} =0,\,\,&{\displaystyle \nabla \times \mathbf {B} ={\frac {1}{c}}\left(4\pi \mathbf {J} +{\frac {\partial \mathbf {E} }{\partial t}}\right)} \label{eq:maxwell1}
\end{eqnarray}
relating the electric field $\mathbf {E}$, magnetic field $\mathbf {B}$, charge  $\rho$, and current  $\mathbf {J}$, and nothing can change those interactions. The equation of motion for a particle with charge $e$, mass $m$, and velocity $\mathbf{v}$ due to the electromagnetic field $\mathbf{E,B}$ and gravitational field $\mathbf{g}$ is given by
\begin{equation}\label{eq:maxforce}
\mathbf{F} = m \frac{d\mathbf{v}}{dt}= e\{ \mathbf{E} + \mathbf{v} \times \mathbf{B}\} + m \mathbf{g}.
\end{equation} 
Equation (\ref{eq:1}) represents the 
solutions that necessarily follow from (\ref{eq:maxwell}-\ref{eq:maxforce}),  
proceeding purposelessly on the basis of the context $C$ (expressed via constraints on the possible values of the variables and control parameters $c$ affecting the form of those constraints) and initial data 
$\textbf{X}(t_1)$. These equations can be written in Lagrangian or 
Hamiltonian form \cite{Arnold}, and are time symmetric and imply energy 
conservation. They have unique solutions as shown for 
example by Arnold (\cite{Arnold}:8) in the Newtonian case and by Hawking and 
Ellis \cite{HE} in the general relativity case \footnote{Some very special 
examples have been devised that do not fulfil the conditions of these 
standard existence and uniqueness theorems, see e.g. \cite{Norton}. 
However  by any reasonable sense of ``randomly''these cases are a set of measure zero and we can ignore them and accept Arnold's uniqueness statement \cite{Arnold}. We thank Tim Maudlin for this comment (and see also \cite{Fletcher}).}   Bifurcations can occur when a small change in a contextual parameter or initial data occurs, but the outcomes are still determined uniquely by the dynamical equations \cite{Arnold}.    
\href{https://en.wikipedia.org/wiki/Statistical_physics}{Statistical physics}  laws  for aggregates of particles follow from the fundamental laws 
\cite{Pen79,Blundell}, which emergent laws  by their 
nature determine probabilistic outcomes $P(q)$ for states 
$q$. They may also have stochastic elements due to  
random environmental effects, leading to stochastical 
dynamics represented by coupling deterministic equations 
of motion to ``noise'' that mimics the effect of many 
unknown variables. Then a stochastic term $\eta(t)$ must 
be added to (\ref{eq:maxforce}) (see \cite{Stochastic}). The 
outcome will then not be determinate, but it will not 
relate in any way to function or purpose.
 
\section{Logic of Life}\label{sec:3}
Life of course obeys the laws of physics, so at each level whatever constraints are implied by physics are obeyed. However additionally living systems behave according to biological logic, leading to what Mayr characterises as goal directed behaviour (\cite{Mayr2004}:52). Life collects and analyses information in order to use it to execute purposeful actions in the light of both genetic heritage and memory (stored information) \cite{Hart99,CamRee05}. This is true from amoeba \cite{Aletal07}  %\footnote{\href{https://www.ncbi.nlm.nih.gov/pmc/articles/PMC2634130/}{Plant neurobiology} is a developing field of study.}
 to all animals \cite{CamRee05} to humans \cite{Kanetal13}. This involves a branching logic where outcomes are chosen on the basis of context, as revealed by the incoming information, and function; this dynamics damps out the effect of fluctuations.  Even plants do something like this, for example in the case of heliotropism  (tracking the sun's motion across the sky) through cell elongation due to the phytohormone  auxin \cite{phototrop}. 

%\subsection{Information usage}
\subsection{Logical branching}
The dynamics followed % Information use %
at each level of biological hierarchies 
 is based on contextually informed logical choices $L$ that tend to support the functions $\alpha$ of a trait $A$ as discussed in section  \ref{sec:function}. That is, in biology action can be actively directed
 (purpose-directed) rather than driven by efficient cause or by chance: 
 \begin{eqnarray}\label{eg:111}
\texttt 
 {Biological dynamics tends to further the function } \alpha\nonumber \texttt{ of a trait } A \\\texttt{ through %the dynamics of
  contextually informed branching logic  } L
\end{eqnarray}
where $L$  is branching logic\footnote{Branching logic is computational logic, see   \cite{AbeSusSus96,ManKim08} for discussion.} of the form 
%is based on contextually informed %logical choices of the form
\begin{equation}\label{eq:2}
\texttt{ 
 L: given context }\, C,\,\texttt{ IF}\,\, T(\textbf{X})\,\, \texttt{THEN}\, F1(\textbf{Y}),\,\texttt{ ELSE}\,\,\textbf{} F2(\textbf{Z}).
\end{equation}Here $\textbf{X}$, $\textbf{Y}$, and $\textbf{Z}$ may be the same or different variables and $T(\textbf{X})$ is the truth value of arbitrary evaluative statements depending on \textbf{X} that can include any combination of Boolean logical operations (AND, OR, NOT, NOR, and so on), possibly 
combined with mathematical operations, while  $F1(\textbf{Y}) $ and $F2(\textbf{Z})$ are %flexible
  outcomes tending to further the function $\alpha$.  Thus they might be 
``If the cat is in, turn the light out, else call the cat in'' (aiming to get the cat in the house) or ``If the calculated future temperature T will be too high and there is no automatic control system, then reduce the fuel flow F manually'' (aiming to keep the temperature in the desired range; a default unstated ``ELSE'' is always to leave the status quo). In the case of flowering plants it might be ``if the sun is shining, open; if not, close''.\\

 The key point is that \textit{the evaluative  function $T(\textbf{X})$ and outcome options
$F1(\textbf{Y}) $ and $F2(\textbf{Z})$ are not determined by
the underlying physical laws, even though they are enabled by them}; they can be shaped jointly by 
evolutionary and developmental processes
\cite{Gil06,GilEpe09} to give highly complex outcomes 
(ranging from phenotype-genotype maps \cite{Wag} to  the 
\href{https://en.wikipedia.org/wiki/Citric_acid_cycle}{citric 
acid cycle} \cite{Krebs,Aletal07} to physiological systems 
\cite{RhoPfl89,CamRee05}), or can be  
\href{https://en.wikipedia.org/wiki/Planning#/media/File:Plann
ing_proces.gif}{planned by human thought} to produce 
desired outcomes \cite{bronowski,Har17}.  One can suggest that trivially any dynamics
of a physical system can be programmed as well in terms of branching logic equivalent to (\ref{eq:2}), so (\ref{eq:2}) is really not different from (\ref{eq:1}), but as discussed in detail in \cite{BinEll18}, physical laws are not the same as programs. Furthermore there is no Hamiltonian or Lagrangian that leads to (\ref{eq:2}),  and in the physics case there is no function $\alpha$ associated with the dynamics, as in (\ref{eg:111}). Physics \textit{per se} is not teleonomic. Furthermore, unlike the case of 
physical laws, where the relevant interactions cannot be 
changed or chosen because they are given by Nature and are 
invariable, these interactions can fulfil widely varying 
biological or social or mental purposes and can be selected for those purposes. It is their
arbitrary nature, essentially similar to Turing's discovery
that a digital computer can carry out arbitrary 
computations, that allows this flexibility, and underlies the two kinds of causation in biology pointed out by Mayr (\cite{Mayr2004}:89-90) because what he labels as a ``biological program'' depends precisely on such logical choices.\footnote{All computer languages have methods of implementing such branching logic.} \\

It is of course not intended here to imply that this kind of causation is determinist: that is why the word ``tends'' is used in (\ref{eg:111}). In particular, chance plays a key role in evolutionary theory \cite{Glymour,Mayr2002} Nevertheless such causation is often reliable \cite{AnimalPhysiology}, for example in the case of the developmental programs referred to by Mayr \cite{Mayr2004}  which underlie developmental biology \cite{Wol02,Gil06}, in the case of molecular machines \cite{Hoff12}, the systems underlying heart function described by Noble \cite{heart}, and the metabolic networks and gene regulatory networks described by Wagner \cite{Wag}.\\

In what follows, we will be looking at the various forms the branching logic (\ref{eq:2}) can take, always taking (\ref{eg:111}) for granted. Biological examples of such logical processes are 
\begin{itemize}
\item
\href{https://www.boundless.com/microbiology/textbooks/boundle
ss-microbiology-textbook/microbial-genetics-7/sensing-and-sign
al-transduction-91/chemotaxis-490-6641/}{Chemotaxis}: 
bacteria detect gradients and move away from poisons and 
towards nutrition \cite{chemotaxis}.
\item \href{https://en.wikipedia.org/wiki/Waggle_dance}{Bee 
dances}: If food is found, bees signal its location  to the
rest of the hive by a waggle dance \cite{beedance}. No dance
implies no food has been found.
\item
\href{http://www.whatisepigenetics.com/what-is-epigenetics/}{
Epigenetics}:  gene expression is controlled 
at lower levels by gene regulatory networks to meet  higher 
level needs \cite{GilEpe09,Nob12,Noble16} (as illustrated in 
Figures \ref{Fig1} and \ref{Fig2}).
\item Human planning of future actions on the basis of 
expected outcomes based on logical choices \cite{Har17}, for 
example computer aided design and manufacture of an aircraft 
on the basis of chosen design criteria \cite{Ell16}. 
\end{itemize}
A particularly crucial form of this branching logic in biology is that implemented in the feedback control circuits that are the foundations of \textit{homeostasis} (\cite{RhoPfl89,CamRee05},\cite{AnimalPhysiology}:8-10). These are of the form (\cite{AnimalPhysiology}:11)) 
\begin{equation}\label{eq:33}
\texttt{IF}\,\, X< X_{MIN}(C)\,\, \texttt{THEN}\,\, X_{INC}(\textbf{Y}),\,\texttt{ ELSE\,IF}\,\, X> X_{MAX}(C)\,\,\, \texttt{THEN}\,\, X_{DEC}(\textbf{Z})
\end{equation}
where $X_{INC}(\textbf{Y})$ is some operation that increases the value of the target variable $X$ through changing the value of the control variable $\textbf{Y}$, and $X_{DEC}(\textbf{Z})$ is some operation that decreases the value of  $X$ through changing the value of $\textbf{Z}$ (which may or may not be the same as \textbf{Y}). The effect is to maintain the value of $X$ to lie between $X_{MIN}(C)$ and $X_{MAX}(C)$; that is the purpose of the interaction. The triggering values $X_{MIN}(C)$ and $X_{MAX}(C)$ are in general dependent on the context (e.g. if the organism is sleeping as against running).  
The default is to leave the situation as is. This is obviously a particular case of (\ref{eq:2}). Control of core body temperature and blood pressure are examples at the macro level, while control of  glucose concentration in  extracellular fluid and of sodium and potassium levels in axons are examples at the micro level. The relevant mathematics is that of \href{https://en.wikipedia.org/wiki/Classical_control_theory}{feedback control systems} \cite{control,Sauro}, using Laplace transforms to model the system and signals, in contrast to the physics equations  (\ref{eq:maxwell}-\ref{eq:maxforce}). Because these homeostatic systems in biology have been tuned through the processes of natural selection to operate successfully, they are in general not subject to the instabilities that can plague feedback control systems in general. Of course a thermostat follows this logic; this is because it has been designed to do so (cf. the comments on machines in section \ref{sec:function}).

\subsection{Phase transitions}\label{sec:phase}
A physicist might suggest that what is proposed here is in 
fact a part of physics  through bifurcations that occur in physics, for example phase 
transitions such as solid/liquid/gas transitions for a 
substance $S$ \cite{Blundell}. These generically have  a form 
like
\begin{eqnarray}\label{eq:22}
\texttt{GIVEN} & \texttt{pressure }P \texttt{ and temperature T},\,\, 
\texttt{IF}\,\, \{P,T\}  \in S_{P,V}\,\, \texttt{THEN S is  solid},\,\nonumber\\&\texttt{ELSE IF }\,\{P,T\}  \in L_{P,V} \texttt{ THEN S is liquid}, \,\,\texttt{ELSE S is gaseous}.	
\end{eqnarray}
Here the context is represented by the pressure $P$ and 
temperature $T$, and $S_{P,V}$, $L_{P,V}$ and $G_{P,V}$ are the 
subsets of the $(P,V)$ plane for solids, liquids, and gases respectively. At first glance this looks like 
it has the form (\ref{eq:2}). However there are two crucial 
differences. First, there is nothing like (\ref{eg:111}) in this case. Second, and related to this, the regions $S_{P,V}$, $L_{P,V}$ and $G_{P,V}$ 
are completely fixed by the physics of the substance S 
independent of history and the environment, and cannot be altered, 
whereas in (\ref{eq:2}), $T(\textbf{X})$, $F1(\textbf{Y})$ 
and $F2(\textbf{Z})$ are contextually set and in some way 
adapted to the environment. As shown by the above examples, 
they have a flexibility that is completely missing in 
(\ref{eq:22}). This is physical determination rather than biological causation (note the contrast with (\ref{eq:33}), where $X_{MIN}(C)$, $X_{MAX}(C)$ are determined through evolutionary processes, see Section \ref{sec:evolution}, as are $X_{INC}$ and $X_{DEC}$).

\subsection{Physical realisation}\label{sec:epigenetic}
\begin{figure}[h]
 \centering
 \includegraphics[width=0.7\linewidth]{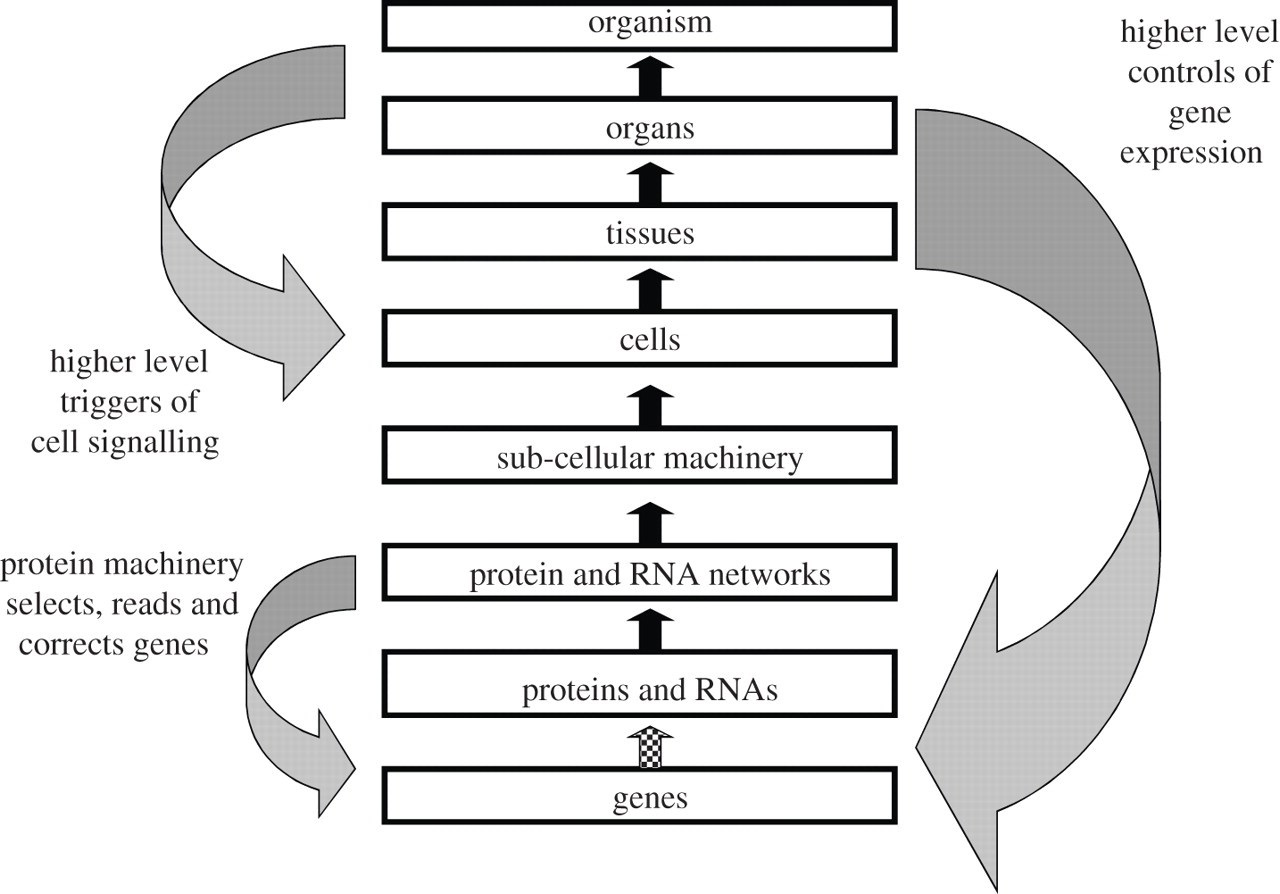}
 \caption{Contextual control \textit{Epigenetic and physiological control of lower level processes by higher level contexts, which determine which logical branching will take place at lower levels by switching genes on and off  on the basis of higher level needs.} From Noble \cite{Nob12}. % (with permission).
 }
 \label{Fig1}
 \end{figure}

 The \href{https://en.wikipedia.org/wiki/Biological_organisation
 }{hierarchy of structure} that underlies existence of life \cite{Ell16}  
 is indicated in Figure \ref{Fig1}.
 \noindent The kind of branching logic indicated in (\ref{eq:2}) occurs at each level in this hierarchy. It occurs at the lower levels via mechanisms such as  voltage and ligand gated ion channels, molecular recognition via lock-and-key mechanisms, and synaptic thresholds, as discussed below.\\
 
\begin{figure}[h]
     \centering
     \includegraphics[width=0.9\linewidth]{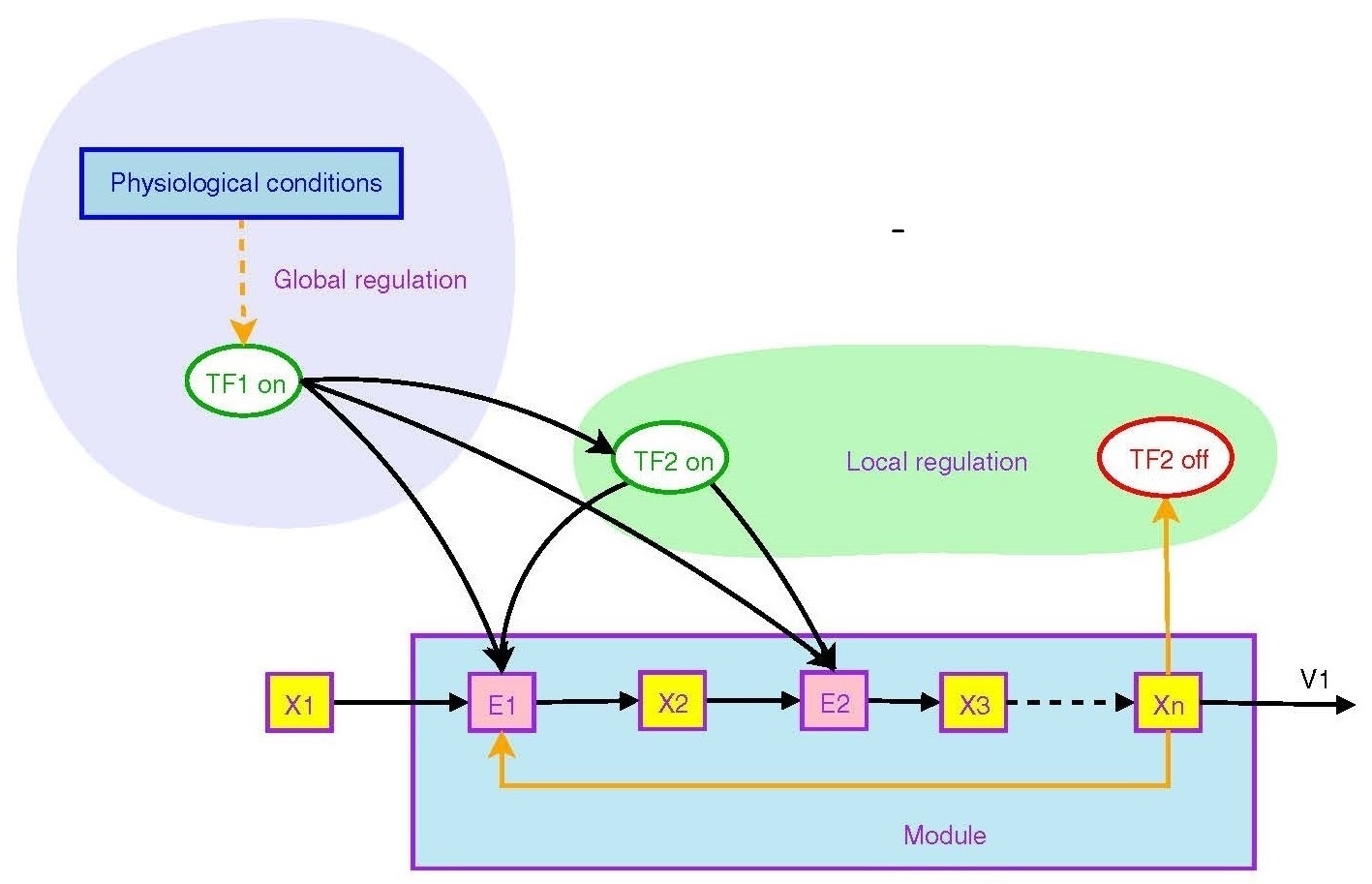}
     \caption{Global and local regulation of a metabolic pathway. \textit{Logical structure of a simple  metabolic pathway module  converting food to energy and protein building blocks, showing some of the many proteins involved, as well as the top-down effect of the large-scale physiological environment. Binding of various factors serves as logic switches.  TF's are transcription factors, E's are enzyme pools, and X's are metabolic pools.} From Goelzer \textit{et al} \cite{Goeetal08} (open access)}.\label{Fig2}
     \end{figure}

 When built into gene regulatory networks (turning genes on and off according to context),  signal transduction networks, metabolic networks, and neural networks, logic gates realised in one of these ways at the lower levels then lead to higher order logical operations such as occur in epigenetic circuits and the functioning of the brain. However the function of the lower level elements is in turn contextually controlled by higher level elements, resulting in contextual emergence \cite{ContextEmerge} where lower level logical choices are set so as to fulfil higher level purpose or function \cite{Nob11,Nob12,Ell16}. Within these circuits exists a unique interaction of constraints and interactions between higher and lower levels within any given system. As Denis Noble described,
 \begin{quote}
 	\textit{``Where do the restraints come from in biological systems? Clearly, the immediate environment of the system is one source of restraint. Proteins are restrained by the cellular architecture (where they are found in or between the membrane and filament systems), cells are restrained by the tissues and organs they find themselves in (by the structure of the tissues and organs and by the intercellular signalling) and all levels are restrained by the external environment. Even these restraints though would not exhaust the list. Organisms are also a product of their evolutionary history, i.e. the interactions with past environments. These restraints are stored in two forms of inheritance - DNA and cellular. The DNA sequences restrict which amino acid sequences can be present in proteins, while the inherited cellular architecture restricts their locations, movements and reactions''.} \cite{Nob11}
 \end{quote}
   Figure \ref{Fig2} shows how logical operations of a lower level module in a metabolic pathway can be regulated by higher order circuits through transcription factors that  initiate and regulate the transcription of genes through the "lock and key" molecular recognition mechanism. They can be ``on'' (able to bind to DNA) or ``off'', thereby controlling the  transcription of genetic information from DNA to messenger RNA.

     The local transcription factor $TF_2$ is sensitive to an intermediate metabolite $X_n$ and modulates the synthesis of enzymes (powerful biological catalysts that control the rate of reactions) in the pathway.  Thus they embody branching logic of the form
          \begin{equation}\label{metabolism1}
           IF\,\, TF_2\,\, on,\,\, THEN\,\, X_2 \rightarrow  X_3,\,\, ELSE \,\,NOT
           \end{equation} 
    which is logic of the form (\ref{eq:2}) operating locally inside the module. However the global regulator $TF_1$, sensitive to higher level variables such as heart rate or blood pressure, can modulate (i) the synthesis of intermediate enzymes, (ii) the synthesis of the local transcription factor $TF_2$, or (iii) both. % with a different logic than at the lower level;
  Thus overall the internal logic of the module acts to produce a ``black box'' whereby conversion of $X_1$ to $X_n$ is controlled by $TF_1$: 
 \begin{equation}\label{metabolism2}
 IF\,\, TF_1\,\, on,\,\, THEN\,\, X_1 \rightarrow X_n,\,\, ELSE \,\,NOT
 \end{equation} 
  This is again a relation of the form (\ref{eq:2}), but at 
   a higher level ($TF_1$ is a higher level variable); the purpose is production of $X_n$ when and only when it is needed. Thus 
   lower level logic circuits such as (\ref{metabolism1}) can 
   be used to build up higher level branching logic such as 
   (\ref{metabolism2}). This is the process of 
   \textit{abstraction} in a modular hierarchy \cite{Boo94}, 
   whereby internal workings are hidden from the external 
   view ($TF_2$, $E_2$, $X_2$, and $X_3$ are local variables 
   whose values are not known to the external world and do 
   not occur in (\ref{metabolism2})). All that matters from 
   the system view is the logic (\ref{metabolism2}) whereby 
   $TF1$ controls conversion of $X_1$ to $X_n$. Note that the global regulation that turns $TF1$ on and off will in general be of the homeostatic form (\ref{eq:33}).  This kind of regulation of lower levels due to higher level 
 conditions can occur between any adjacent level in the 
 hierarchy of structure and function\footnote{This is like 
 the way there is a tower of virtual machines in a digital 
 computer, with a different formal logic operational at each 
 level \cite{Ell16}, and parameter passing from the main program to submodules.} and through it, metabolic regulation can have  a major effect on gene expression \cite{metabolic16}. 	\\

 This 
     process of \textit{black boxing} 
     (\cite{Asbycybernetics}:\S6) of lower level logic to 
     produce higher level logic in biology contrasts strongly 
     with the process of \textit{coarse graining} \cite{Pen79} 
     to produce higher level variables and effective laws out 
     of lower level variables and effective laws in physics. \\
     
 \textit{Multiple realisation} of higher level functions and processes at lower levels is a key feature in the emergence of complexity such as life, and is central to the idea of black boxing. The key analytic idea  is that of identifying \textit{functional \href{https://en.wikipedia.org/wiki/Equivalence_class}{equivalence classes}} \cite{Aletal07,Ell16}: each equivalence class is a set of lower level properties that all correspond to the same higher level structure or function. Equivalence classes at a lower level collect elements whose differences are irrelevant for the emergent target feature at the higher level.
 	This degeneracy occurs in all biology in relation to the underlying microbiology and physics. An example is the way developmental systems are related to the genome: a vast number of different genomes (a \textit{genotype network}) can create the same phenotype \cite{Wag};	any one
 	of them can be selected for and will do the job needed. This huge degeneracy solves the problem of how biologically effective alternatives can be explored in the time available since the start of life, as explained by Wagner \cite{Wag} in his important book. %, by exploring the relevant possibility spaces via adaptive selection.
 	%It is the equivalence class at the lower level that  realises a high Equivalence classes are key to the downward causation that enables true complexity to emerge \cite{Ell16}.
 	%All of these lower level states correspond to the samehigher level state; entropy is a measure of how many lower level states correspond to a specific higher level state \cite{Sto15}. 
 	These are what get selected for when adaptation takes place; and it is the huge size of these equivalence classes that enables adaptive selection to search out the needed biomolecules on geological timescales    \cite{Wag}. Whenever you can identify existence of such functional equivalence classes, that is an indication that top-down causation is taking place \cite{AulEllJae08}.\footnote{See also the quote by Weiss in Section \ref{sec:link}.
 	} %\footnote{G. Auletta, G. Ellis, and L. Jaeger (2008) "Top-Down Causation: From a Philosophical Problem to a Scientific Research Program"  \href{http://rsif.royalsocietypublishing.org/content/5/27/1159}{\textit{J R Soc Interface} B: 1159-1172}.} 
 	Such multiple realisation occurs in cases such as the metabolic networks in a cell and gene regulatory networks: many different lower level realisations can occur of the needed higher level functions.\\

  All these logical operations are based 
 in physics, but are quite different than the logic of 
 physical laws \textit{per se}. How are they realised through 
 the underlying physical stratum?

\section{The Physical Basis: Linking Physics and Biology}\label{sec:4}
To give the discussion a specific context, we will now focus on the brain.
 
\subsection{The nervous system}

Brains are based in the underlying physics through the 
operation of neurons linked by synapses and structured in 
neural networks, in particular  forming layered columns in 
the neocortex \cite{Kanetal13}. Neurons receive  spike trains via dendrites, which flow to the 
nucleus where a summation operation is performed, and 
resulting spike trains then flow down axons to synapses where 
a further summation process takes place; signals are passed 
to other neurons if the outcome is above a threshold 
\cite{Kanetal13}. The behaviour of currents in dendrites and 
axons is governed by the underlying physics (described by classical 
equations (\ref{eq:maxwell}-\ref{eq:maxforce})).\footnote{Essentially quantum mechanical interactions probably play no significant role in this functioning,  %based on the electromagnetic force 
	but they underly the existence of the molecular structures of neurons.} The purpose is to underlie the functioning of the nervous system that enables an animal to anticipate and counter threats to its existence, thus enhancing its chances of survival.
\\

 The 
\href{https://en.wikipedia.org/wiki/Hodgkin-Huxley_model}{Hodgkin-Huxley equations} \cite{HodgHux} charactize ion and 
electron flows that underlie existence of action potential 
spike trains in neurons (\cite{AnimalPhysiology}:132-1139). These equations (see Appendix \ref{sec: hodg_Hux} for details) follow from the 
structure of axons and dendrites, and in particular from the 
the existence of ion channels (\cite{Cat00}, \cite{AnimalPhysiology}:141-150)  that allow ions to selectively 
flow in and out of the cell membranes. The constants in these 
equations are not universal constants, but are constants 
characterising the axon structure and environment. They cannot be deduced from the laws of physics alone  \cite{Scott} (see Appendix \ref{sec: hodg_Hux}).

\subsection{The link of physics to logic: the molecular basis}\label{sec:link}
 
\begin{figure}[h]
 	\centering
 	\begin{minipage}{0.45\textwidth}
 		\centering
 	\includegraphics[width=0.9\linewidth]{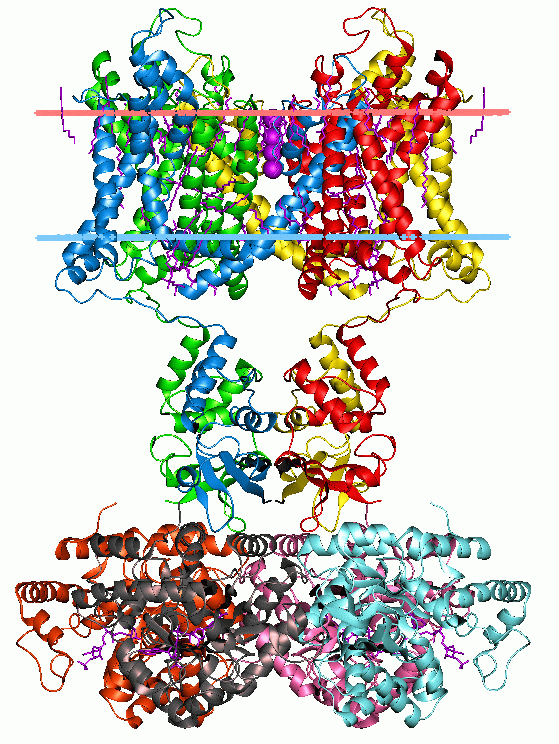}
 	\caption{Potassium ion channel structure in a membrane-like environment. \textit{This 3-dimensional structure alters according to the voltage difference across the membrane, hence allowing or impeding ion passage.}
 	%Calculated hydrocarbon boundaries of the lipid bilayer are indicated by red (top) and blue (lower) dots.} 
 	Diagram by Andrei Lomize. From  \href{https://en.wikipedia.org/wiki/Voltage-gated_potassium_channel}{Wikipedia}.}  
 	\label{fig:voltagegated}
 	\quad 
\end{minipage}\hfill
\begin{minipage}{0.45\textwidth}
\centering
	\includegraphics[width=0.9\linewidth]{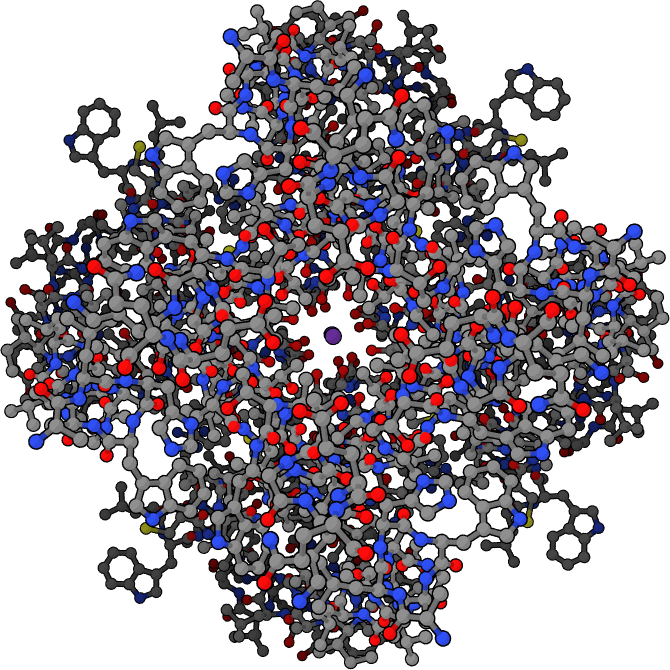}
	\caption{Potassium ion channel functioning \textit{Top view of potassium ion (purple, at centre) moving through potassium channel when channel is open.} From  \href{https://en.wikipedia.org/wiki/Potassium_channel}{Wikipedia}}
	\label{fig:potassiumchannel1}
\end{minipage}
\end{figure}

 The logical structure of what happens is enabled by particular proteins: namely 
 \href{https://en.wikipedia.org/wiki/Depolarizing_pre-pulse}{voltage gated ion channels} in axon and dendrite membranes   (\cite{Cat00,Mag17},\cite{AnimalPhysiology}:146-151) (see Figs \ref{fig:voltagegated} and \ref{fig:potassiumchannel1}). 
 They lead to controlled flow of sodium, potassium and 
 chloride ions into and out of the axons and dendrites, leading to action potential spike chain  propagation along the axons and dendrites. Their molecular structure and function is discussed in (\cite{AnimalPhysiology}:139-147).
  \noindent The 
 ion channels implement branching logical operations with the following 
structure:\footnote{In practice, the response function is not 
 discontinuous as in this representation, but is a smoother 
 curve such as a logistic curve linking `on' and `off' states. The principle remains the same: but one now uses fuzzy logic \cite{fuzzy}.}
  \begin{equation}\label{eq:3}
 \texttt{IF voltage}\,\, V > V_0 \,\,\texttt{THEN let ions flow, ELSE not}
 \end{equation}
 which is of the form (\ref{eq:2}). % They  also occur with the inequality reversed.
The purpose is to facilitate the propagation of action potentials in axons, and so enable functioning of the nervous system \cite{AnimalPhysiology}. It is the structural form of the 
 \href{https://en.wikipedia.org/wiki/Voltage-gated_ion_channel#/media/File:Open_and_closed_conformations_of_ion_channels.png}{ion channels} - the detailed three-dimensional molecular 
 configurations characterised as 
 \href{https://en.wikipedia.org/wiki/Protein_tertiary_structure}{tertiary} and 
 \href{https://en.wikipedia.org/wiki/Protein_quaternary_structure}{quaternary} structures (see Figs \ref{fig:voltagegated} and \ref{fig:potassiumchannel1})   that enables conformational change in response to local conditions that controls the flow of ions in and out of the cell wall, and so enables this branching logic to emerge out of physics %in the brain
      at the lower levels. This then underlies emerging branching logic at higher levels in the cortical networks in the brain  \cite{Kanetal13,beimGraben},  effectively through predictive coding processes \cite{predictive} enabled by those networks. Thus the relevant structure enabling emergence of branching logic is 
  that of  proteins \cite{PetRin09}  imbedded in the cell wall. %: see \href{https://www.bnl.gov/bnlweb/history/nobel/images/potassium_channels-264px.gif}{here} 
   %and \href{http://lab.rockefeller.edu/mackinnon/assets/image/XTFigure.png}{here} 
  %for the  structure of one particular family of these channels. 
    Given the existence of the ion channels with their specific 3-dimensional form, Maxwell's equations (\ref{eq:maxwell},\ref{eq:maxwell1}) together with the equations of motion for particles (\ref{eq:maxforce}) underlie what happens by controlling the flow of ions.
  The flow of ions through the ion channels is governed by these physical laws and so has the physics form (\ref{eq:1}). \\
  
  The implication is that, at least in the brain, 
\begin{quote}
\textbf{Molecular basis of ``IF ... THEN ...  ELSE ...." branching logic  in biology}: \textit{Biomolecules perform branching logical
operations because of their tertiary and quaternary structure}.
\end{quote}
%This applies equally in many other biological contexts (see Section \ref{sec:more_general} below), where t
Once physical implementation of logical processes have been achieved at the lower levels, this provides the building blocks for implementing logical processes at higher levels.  Basic logical units can be used to give the basic operations AND, OR, NOT, and can then can then be  combined via specific neuronal connections and weights in neural networks, 
% However, it is
in an almost incomprehensible way %how
 with thousands of synaptic inputs on each neuron, %could lead
 leading to the coordinated actions involved in learning, memory, and higher level cognitive function. The key point here is that the micro level logical operations given by the biomolecules lead to contextual emergence of higher level operations \cite{ContextEmerge} because they themselves are contextually controlled by higher level functions, through being incorporated in neural networks engaged in higher level logical operations \cite{beimGraben,Ell16}. \\

 %However,
 More complex logical operations occur, for example N-methyl-D-aspartate (NMDA) receptors \cite{NMDA} are one %such
  class of receptors capable of responding to the high density of synaptic inputs on a single neuron. These receptors have a characteristic % seven transmembrane domain coupled to G proteins 
 heterotetramer between two NR1 and NR2 subunits that mediate numerous biological effects within neurons \cite{Nestetal15}. Most notable of these effects relate to learning and memory, particularly with respect to long-term potentiation (LTP) and synaptic plasticity in the hippocampus. Unlike typical voltage-gated ion channels previously mentioned, NMDA receptors respond to the coordinated input from many synapses, leading to global depolarization across the cell membrane before activating and allowing calcium entry into the cell \cite{Nestetal15}. Thus, NMDA receptors are linked to higher level functions in humans. The integrative nature of NMDA receptors illustrates the holistic nature of biological systems expressed by the late developmental biologist, Paul A. Weiss. In his book, \textit{The Science of Life}, Weiss wrote \cite{Wei78}, 
 \begin{quote}
 	``\textit{The complex [biological organism] is a system if the variance of the features of the whole collective is significantly less than the sum of variances of its constituents; or, written in a formula:} 
 \begin{equation}
 V_s %\leq
 <(V_a+V_b +...+V_n).
 \end{equation}
\textit{ In short, the basic characteristic of a system is its essential invariance beyond the much more variant flux and fluctuations of its elements or constituents... This is exactly the opposite of a machine, in which the pattern of the product is simply the terminal end of a chain of rigorously predefined sequential operations of parts. In a system, the structure of the whole coordinates the play of the parts; in the machine, the operation of the parts determines the outcome. Of course, even the machine owes the coordinated functional arrangement of its parts, in last analysis, to a systems operation - that of the brain of its designer}'' .
 \end{quote} Similarly, there exists a dynamic interplay between single voltage-channel receptors and receptors responding to the global interplay of all synaptic inputs on postsynaptic neurons. Thus, the brain exists as a dynamic entity of intertwining top-down and bottom-up processes \cite{Ell16,SethFreeEnergy}.
\\

 Energy is of course used in carrying out these logical processes (see Appendix \ref{sec:energy} for energy processes in the brain), and local energy minimisation will occur as part of what is going on (e.g. in the protein folding that converts one-dimensional strings of amino acids to the very complex tertiary and quaternary structures of proteins \cite{PetRin09}). But energy or entropy considerations  will not by themselves produce  the desired logical structures and operations that enable the function and purpose characteristic of life \cite{Hart99} to emerge. %; in particular, functioning of the brain is very expensive in energy terms \cite{Godrfysmith}.
  Actually the necessary energy usage for cellular function is controlled by complex metabolic regulatory networks \cite{Wat13} that determine what energy transactions; Figure \ref{Fig2} gives a simple example, showing how they rely on basic branching logical operations (\ref{eq:2}) mediated by transcription factors that can be ``on'' or ``of''. Thermodynamics and statistical mechanics are not by themselves enough to capture what is going on. \\
 
 \noindent In summary, \textit{\textbf{Given the right cellular context \cite{Hoff2017}, biomolecules such as ion channels \cite{Cat00,Mag17} can be used to make logic gates, and the higher level physiological systems in which they are imbedded \cite{CamRee05,Goeetal08,RhoPfl89,Kanetal13} enable emergence of complex life processes and branching logic}}.\\
 
\noindent Similar issues arise at synapses \cite{Kanetal13}, where a branching logic
  \begin{equation}\label{eq:3}
 \texttt{IF summed input voltage}\,\, V > V_0 \,\,\texttt{THEN fire action potential, ELSE not}
 \end{equation}
 holds, enabled by voltage-gated $Ca^{++}$ channels in conjunction with pre- and post-synaptic neurotransmitter transporters, and post-synaptic receptors.
 
 \subsection{More general biological contexts}\label{sec:more_general}
\begin{figure}[h]
 	\centering
 	\includegraphics[width=0.8\linewidth]{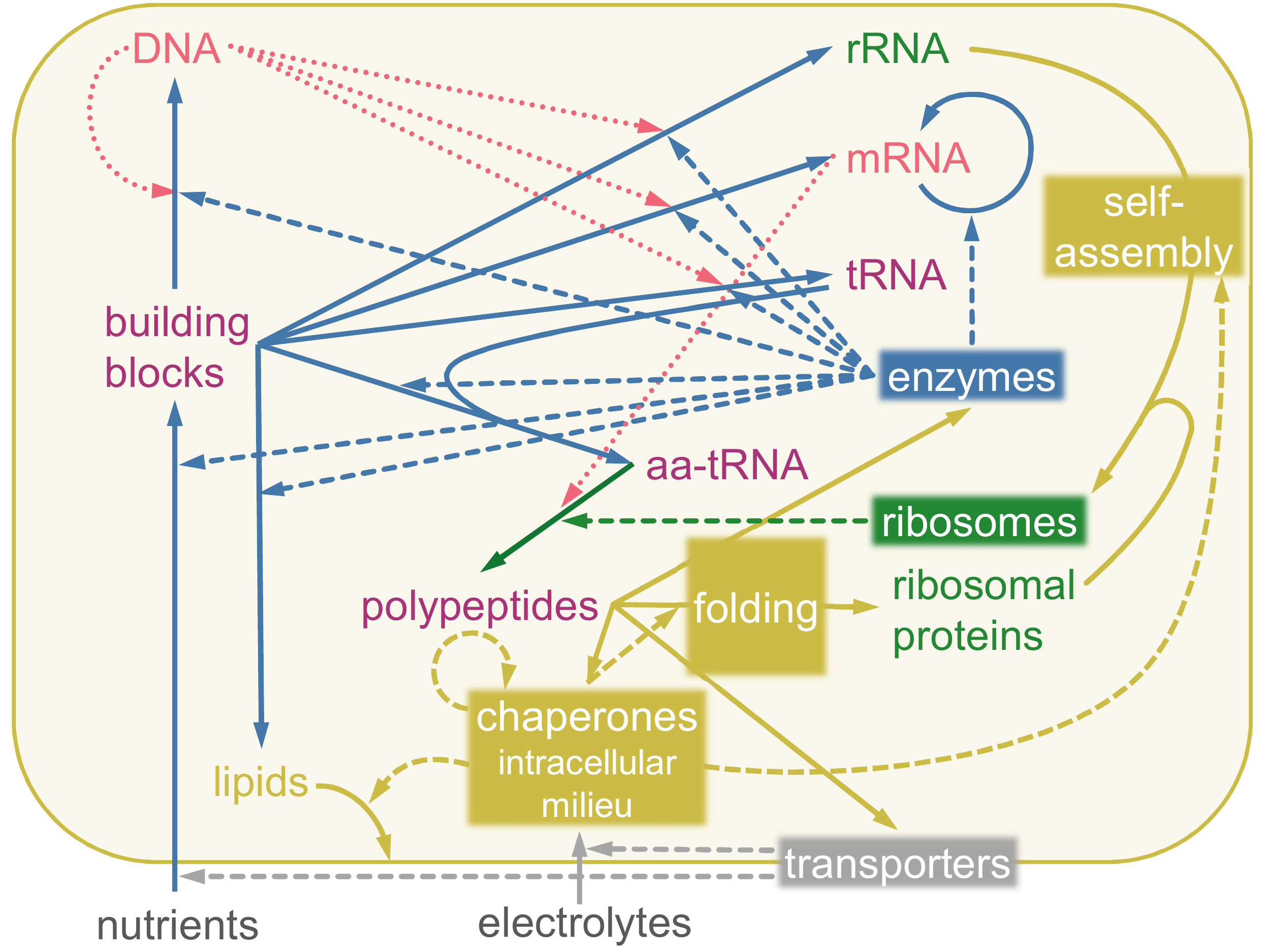}
 	\caption{The structure of cell interactions outlined \textit{The self-fabricating organization of the cell as a superposition of enzyme- and ribosome-catalysed covalent chemistry (blue), supramolecular folding and self-assembly (yellow), and membrane-associated electrolyte transport (grey).} Figure from Hofmeyr \cite{Hoff2017}, with permission.} 
 	\label{fig:hof17cell}
 \end{figure}

 The basic branching logic discussed here more generally underlies the functioning of all cells, as illuminatingly discussed by Hofmeyr \cite{Hoff2017,Hoff2017a}. There are three major aspects: metabolic, genetic (the reading of the DNA code), and epigenetic. The latter \cite{GilEpe09,Noble16} was discussed to some degree in Section \ref{sec:epigenetic} (and see \cite{GilEpe09,Noble16}). We will discuss the other two in turn here  (see Figure \ref{fig:hof17cell} for the context, which controls what happens).
 \\

\textbf{Metabolism} \cite{Hoff2017} The purpose of metabolism is to produce molecules and free energy needed by the cell in usable form, which are crucial for its function and survival. 
  There are three key elements here:
 \begin{enumerate}
 	\item \textit{Enzymes and ribosomes catalyse covalent metabolic chemistry}, providing the  building blocks of life. This is only possible because of the presence of  efficient catalysts, particularly enzymes,  that are highly specific with respect to the substrates
 	they recognize and the reactions they catalyze. The branching logic is (cf. Section \ref{sec:epigenetic})
 	 \begin{equation}\label{eq:4}
 	\texttt{IF catalyst for reaction R1 present} \,\,\texttt{THEN  R1 proceeds, ELSE not}
 	\end{equation}
		Its molecular basis is the relevant lock and key recognition mechanism \cite{Aletal07}.
 	 	\item \textit{The intracellular milieu enables supermolecular chemistry processes} such as folding and self-assembly of polypeptides and nucleic acids into functional catalysts and transporters. This requires chaperones that guide folding in a conducive
 	 	intracellular context comprised of water with a homeostatically maintained electrolyte composition different from that of the external environment: ``G\textit{iven a watery environment with a specific pH, temperature,
 	 	ionic strength, and electrolyte composition, the folding of polypeptides into their
 	 	functional state is a spontaneous process that is determined by their amino acid sequences (primary structures)}'' \cite{Hoff2017}. Overall, 
  	\begin{equation}\label{eq:351}
  		\texttt{IF milieu suitable THEN folding takes place, ELSE not}
  	\end{equation}This requires a cell membrane, which, as in the case of axons discussed above, contains proteins that selectively transport in and out 
 	 	nutrients and electrolytes.
 	\item \textit{Membrane transport maintains the intracellular milieu as required by 2.}, in particular its electrolyte composition. The branching logic is essentially the same as (\ref{eq:3}):
 	\begin{equation}\label{eq:35}
 	\texttt{IF concentration}\,\, C < C_0 \,\,\texttt{THEN let ions flow in, ELSE not}
 	\end{equation} 
 	or else the same with $>$ replaced by $<$. Indeed this is a special case of the homeostasis relation (\ref{eq:33}).
 \end{enumerate}
In the case of synapses, this all occurs in a complex environment of blood vessels (capillaries) and glial cells (astrocytes) that control the availability of ATP for neuronal metabolism (see \cite{Russell} and Appendix \ref{sec:energy}).\\

\textbf{Reading the genetic code} \cite{Aletal07,Wat13} The purpose of the genetic code is to specify the sequence of amino acids that will lead to existence of proteins with crucial biological functions. Following Barbieri \cite{Barbieri}, Hofmeyr defines a code as follows \cite{Hoff2017a}: \begin{quote}
	``\textit{A code is a small
set of arbitrary rules selected from a potentially unlimited number
in order to ensure a specific correspondence between two
independent worlds.}''
\end{quote}
Thus one considers two finite sets $S$ and $T$,  the source and target alphabets, the code being a mapping
$C : S \rightarrowtail T^*$  that maps each symbol in $S$ to a sequence of
symbols (words) in $T^*$, the set of all sequences over the alphabet $T$. The extension of $C$ is a mapping
$W : S^* \rightarrowtail T^*$ that uniquely translates sequences of source symbols
into sequences of target symbols. Hofmeyr \cite{Hoff2017a} then shows how this definition applies to the Morse code and the genetic code. The latter is a mapping $C_G$  from the source alphabet of
64 triplet sequences (codons) formed from the four nucleotides
in mRNA (A, G, C, U) to the target alphabet of the twenty
amino acids found in proteins plus the stop signs:
\begin{equation}\label{eq:genetic_code}
C_G = \{GGU \rightarrowtail Gly;GGC \rightarrowtail Gly;GCU \rightarrowtail Ala;UUA \rightarrowtail stop; ...\}
\end{equation}
The extension $C_G^*$ of the genetic code translates sequences of codons
into sequences of amino acids. Now (\ref{eq:genetic_code}) is just a special case of (\ref{eq:2}): given the cellular context ${\cal C}$ (without which no reading of the genetic code would take place), the branching logic is 
\begin{equation}\label{eq:2222}
\texttt{IF triplet GGU THEN Gly ELSEIF triplet GGC THEN Gly ELSEIF ...},	
\end{equation}
 with a unique mapping specified for each of the 64 triplets.  The purpose of this process is to produce the proteins that are needed in order that cellular life can proceed as needed for a viable organism to exist. This particular highly degenerate mapping \cite{Wat13,Wag} implemented by cellular processes \cite{Aletal07} has been determined by the specific historical events of the evolutionary history of life on Earth \cite{CamRee05,Godrfysmith}: many other mappings are chemically possible. Physics by itself does not determine the specific mapping (\ref{eq:genetic_code}) or logic (\ref{eq:2222}). 
 
\section{Existence of the Relevant Proteins}\label{sec:5}
Two issues arise here: the possibility of  existence of the biomolecules needed, for example those that comprise ion channels, and how they come into being.

\subsection{Their possible existence}\label{sec:exist}
The possible existence of biomolecules, and particularly the proteins that govern biological activity \cite{PetRin09}, 
results from quantum interactions mediated by the electromagnetic force \cite{Wat13}. Indeed both the existence of atomic nuclei and of molecular  binding forces cannot be explained classically. But given the nature of physics as we know it (with particular values for the fundamental constants \cite{Uzan}), the nature of everyday scale structures is controlled by electromagnetism together with quantum physics. There is a resulting space of possible proteins \cite{PetRin09} of vast dimensions: an unchanging space of all possible molecular structures  \cite{Wag}. Such a platonic space has been proposed to explain the common structural motifs found within all proteins in biological systems %Michael Denton 2009 
\cite{DenMarLeg02}.

\subsection{Their coming into being}\label{sec:evolution}
Given this possibility space, how have the specific proteins that exist and control biological function come into being?  
This has developmental and evolutionary aspects. \\

\textbf{Developmental aspects} The relevant proteins exist 
because of the reading of the genetic information written 
into our DNA \cite{Aletal07,Wat13} through molecular processes, resulting in amino 
acid chains that fold to form biologically active proteins. 
This reading of the genotype takes place in a contextual way 
\cite{Wol02,Gil06,Nob12} because of epigenetic processes \cite{GilEpe09} taking place in a developmental 
context \cite{Oyaetal01}. 
\\

\textbf{Evolutionary aspects} How did that genetic 
information came to exist? Equivalently, how did the specific 
proteins that actually exist \cite{PetRin09} come be selected 
from all of those that might possibly have existed, as 
characterised by the vast space of protein possibilities 
\cite{Wag}?  These extraordinary complex molecules with 
specific biological functions (for example, 
\href{https://en.wikipedia.org/wiki/Hemoglobin}{hemoglobin} 
exists in order to transport oxygen in our blood stream; chlorophyll exists in order to enable plants to harvest solar energy) 
cannot possibly have come into being simply through  random bottom-up self assembly, or by  genetic mutation, drift, recombination, or migration alone (\cite{MorrisLundberg2011}:21), \textit{inter alia} because there are so many proteins each performing a different physiological function \cite{PetRin09}. They have to have been selected for
through the process of Darwinian 
\href{https://en.wikipedia.org/wiki/Natural_selection}{adaptive selection} \cite{Darwin,Mayr2002,CamRee05,MorrisLundberg2011,Kampourakis} occurring at the organism level, with these selective outcomes chaining down to the genotyope level \cite{Ell18} within a functional cellular context  \cite{Hoff2017}.   Massive degeneracy in the 
genotype-phenotype map plays a crucial role in enabling new 
genotypes to come into being in the time available 
\cite{Wag11}; and doing so in such a way that the organism remains viable at each step.\footnote{The idea that natural selection increases fitness in populations is queried by population geneticists. See \cite{Ell18} for an in depth response that supports the view just stated.} \\

 Thus we are broadly supporting the view labelled as ``Explanatory Adaptationism'' by Godfrey-Smith \cite{Godfer-Smith2001}. In discussing various views, he states ``\textit{I will set
 aside the issue of whether variation and selection really do adequately explain apparent
 design; I assume that they do, in accordance  with standard Darwinian ideas}.'' This is all we need for the purposes of present paper. It is also described by West and Gardner as follows \cite{WestGardner}: 
\begin{quote}
	``\textit{The most striking fact about living organisms is the extent to
 which they appear designed or adapted for the environments
 in which they live. The theory of natural selection
 provides an explanation. Darwin \cite{Darwin} pointed out that those
 heritable characters that are associated with greater reproductive
 success will tend to accumulate in biological populations,
 and he argued that this will lead organisms to appear
 as if they were designed to maximise their reproductive success.
 Hence, natural selection explains the appearance of
 design without invoking an intelligent designer \cite{MorrisLundberg2011}.
 More generally, the currently accepted paradigm for the
 study of adaptation, in fields such as animal behaviour,
 evolutionary ecology and sociobiology, is that organisms
 should appear designed to maximise their inclusive fitness,
 rather than their reproductive success. Inclusive fitness
 captures how individuals are able to influence the transmission
 of their genes to future generations - they can either
 influence their own reproductive success (direct fitness) or
 the reproductive success of other individuals with which
 they share genes (indirect fitness).}''
\end{quote} 
 
 This adaptive selection process is discussed illuminatingly by Morris and Lundberg \cite{MorrisLundberg2011}, who summarise it as follows:
 \begin{equation}\label{eq:selectdyn}
 \texttt{genes }\Rightarrow \texttt{ traits } \Rightarrow \texttt{ function }\Rightarrow \texttt{ fitness }\Rightarrow \texttt{adaptive value }\Rightarrow \texttt{genes}
 \end{equation} and so on in cyclic fashion, with each step in (\ref{eq:selectdyn}) in fact being very complex. 
 This overall process results in information about the environment being embodied in the sequences of base pairs 
in the DNA \cite{Wat13}  %this process writes information  into genes 
\cite{Sto15} (for example the genome of a polar bear causes its fur to be white, implicitly coding the fact that the polar environment is white (\cite{Kampourakis}):177). Then reading out that information by cellular 
processes \cite{Wat13} creates the string of amino acids that 
forms proteins such as  haemoglobin. It then has to 
\href{https://en.wikipedia.org/wiki/Protein_folding}{fold} to 
give its biologically active form. In principle that step is 
an energy minimisation operation; in practice it requires 
\href{https://en.wikipedia.org/wiki/Chaperone_(protein)}{molecular chaperones} to achieve the required folding \cite{Hoff2017}.  These are 
a further set of proteins that have to be coded for in DNA and 
whose existence has to be explained on the basis of natural 
selection. Additional energy is also used for 
post-translational modification to produce the final protein 
structure. In this context, and only in this context, self-assembly can take place \cite{Hoff2017}.%Additionally, growingevidence suggests that epigenetic mechanisms, such as DNA methylation, may also be involved in rapid adaptation to new environments \cite{PigMul00,epigenetic}.

\subsection{The basic selection process}\label{sec:selection}
\begin{figure}[h]

Darwinian selection is a special case of the  generic selection process that is ubiquitous in biology. The basic generic selection process is that a random input ensemble of entities is filtered to produce an ordered output ensemble fulfilling some environmentally dependent selection criteria (Figure \ref{fig:thebasicselectionprocess}).

\centering
\includegraphics[width=0.7\linewidth]{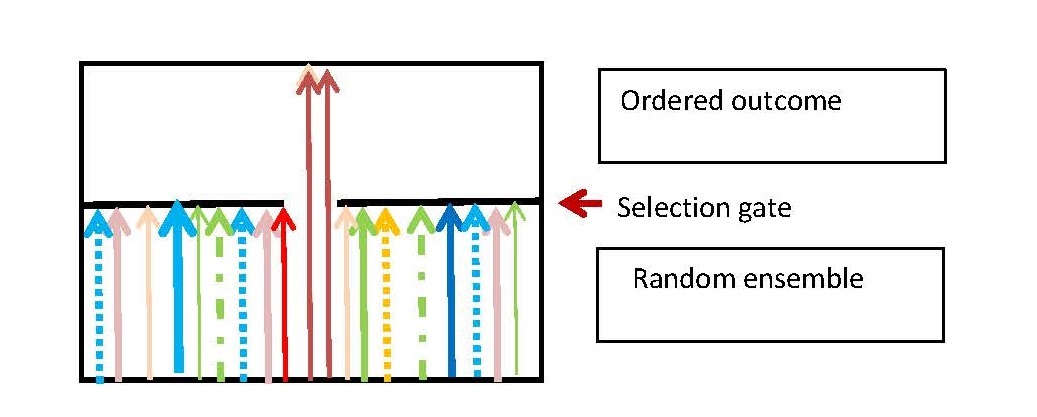}
\caption{\textit{The basic selection process}}
\label{fig:thebasicselectionprocess}
\end{figure}
%Example: Maxwell?s demon
The branching logic of the process is a special case of that given in equation (\ref{eq:2}):
\begin{equation}\label{eq:select}
\Pi_S(X): \{\texttt{IF}\,\, X \notin S(C,{\cal{E}})\,\, \texttt{THEN delete}\,\, X\}	
\end{equation}
where $S$ is the subset of elements selected to survive if $C$ is the selection criterion and $\cal{E}$ the environmental context. 
The effect on the ensemble $\{E(X)\}$ is a projection operation:
\begin{equation}\label{eq:project1}
\Pi_S:  \{E(X)\} \rightarrow \{\hat{E}(X): X \in S(C,{\cal{E}})\}.
\end{equation}
So now projecting again, $\Pi_S$ leaves the new ensemble $\{\hat{E}(X)\}$ invariant:
\begin{equation}\label{eq:project_id}
\Pi_S: \{\hat{E}(X)\} \rightarrow \{\hat{E}(X)\}.
\end{equation}
The purpose of the process is to produce a population of entities that fulfil the selection criterion $C$. 
A simple example in the logical case is deleting 
emails or files on a computer according to some 
criterion $C$; the basic physics case is 
\href{https://en.wikipedia.org/wiki/Maxwell's_demon}
{Maxwell's Demon} \cite{Maxwell}, where the 
criterion $C$ for allowing a molecule to pass the 
trapdoor  is $|{\textbf v}|>v_0$ where $|{\textbf 
v}|$ is molecular speed; a biological case is the immune system, deleting invading pathogens \cite{AnimalPhysiology,RhoPfl89}. \\

Darwinian selection   
\cite{CamRee05,Godfer-Smith2001,Mayr2002}  has the structure overall 
(\ref{eq:select}) where $C$ is a measure of 
inclusive fitness \cite{WestGardner} in the context of the 
environment, and the input ensemble at each time 
$t_2$ is a randomised variant of the 
output of the previous process at previous time 
$t_1$: %thus %, randomised to some degree: 
\begin{equation}\label{eq:randomise}
\{E(X)\}(t_2) = R\{\hat{E}(X)(t_1)\}.
\end{equation}
%\begin{figure}
%\centering
%\includegraphics[width=0.7\linewidth]{selection}
%\caption{Darwinian selection}
%\label{fig:selection}
%\end{figure}
Here $R$ is a randomisation operation based 
in mutations, genetic drift, recombination, and 
horizontal gene transfer. Thus the process is a 
continually repeated 2-step process (\cite{Mayr2002}:130-133): variation 
(\ref{eq:randomise}), which is where differential reproductive 
success enters, followed by elimination 
(\ref{eq:select}), which is where differential survival rates matter (this only 
requires selection of individuals who are ``good enough'' 
(\cite{Mayr2002}:130-131); they don't have to be the 
fittest)\footnote{The word ``fittest'' does not occur in the 
index of Darwin's book \cite{Darwin}, Mayr's book 
\cite{Mayr2002}, Kampourakis' book \cite{Kampourakis}, or the 
book by Morris and Lundberg \cite{MorrisLundberg2011}.}. It is 
the elimination 
phase (\ref{eq:select}) that leads on average, in 
suitable 
circumstances,\footnote{The notion 
that natural selection is a process of average fitness 
increase in a population is queried in population genetics 
\cite{Bir2015}, even though it is common in 
writings of physiologists and many evolutionary biologists (as quoted 
above). A referee writes ``adaptation does not in general tend to increase the mean fitness of a population, and neither therefore does selection leading to adaptation - here, Fisher's fundamental theorem  suffices.'' That theorem states the change in average inclusive fitness ascribed to
the action of natural selection is equal to the genetic variance in inclusive fitness \cite{Gardner,Fisher}. Birch  \cite{Bir2015} suggests the issue has to be studied 
on a case by case basis, and one 
might suspect that selection of proteins of the 
kind discussed here would lie in the class where 
selection does indeed lead  to increase in fitness on  average in a population %and hence adaptation,
 because 
there (a) is no obvious mechanism whereby their 
selection occurs at a cost to other individuals in the population, and (b) otherwise we cannot account for their adaptedness to the job they do in a physiological context \cite{Wag11}. We 
will from now on take this to be the case in this 
paper, but acknowledge it needs further study. See \cite{Ell18} for steps in this regard.}  
to selection of individuals with traits that are 
better fit to the environment. % .(while it may lead to 
%an increase or decrease in the genetic variation in 
%populations of organisms). 
The combination of these two processes leads to inclusive fitness \cite{WestGardner}.   
Thus this adaptive selection process \cite{MorrisLundberg2011} functions to produce individuals fit to survive in a specific environmental context through their physiology and functioning even though the process has no intentional ``purpose'' (\cite{Mayr2004}:58).  
It is the remorseless continual repetition of the 
process of variation (restricted by physical, physiological, 
and development possibilities \cite{Nob11}) and 
subsequent selection that gives evolution its 
extraordinary creative power, underlying the 
emergence of complex life forms 
\cite{Darwin,Mayr2002,CamRee05}. While much of this 
happens by alterations in DNA, additionally, 
growing evidence suggests that epigenetic 
mechanisms, such as DNA methylation, may also be 
involved in rapid adaptation to new environments 
\cite{PigMul00,epigenetic}.
%This is all constrained by physics: in essence, physical laws 
%themselves provide an important contribution for direction and 
%function of living systems.
\\

 Given the caveat mentioned in Footnote 12, the overall process results in adaptation of organisms to the environment (Mayr \cite{Mayr2002}:166-169; Kampourakis \cite{Kampourakis}:172-184).  %Here we accept the definitions given by  Dobzhansky \cite{Dobzhansky}:
 %1. Adaptation is the evolutionary process whereby an organism becomes better able to live in its habitat or habitats.
 %2. Adaptedness is the state of being adapted: the degree to which an organism is able to live and reproduce in a given set of habitats.
 %3. An adaptive trait is an aspect of the developmental pattern of the organism which enables or enhances the probability of that organism surviving and reproducing. %Thus  (cf. Kampourakis (\cite{Kampourakis}:173--178)) we adopt the historical definition of adaptation.
  There can be selection for adaptive characters, or selection against organisms that do not have such characters (\cite{Kampourakis}:179--184).
 %; equivalently, the organism has adapted so as to reduce surprisal \cite{Sto15}  s it moves through its environment.
 % Adaptation takes place either by the organism adapting its structure and behaviour to the environment, or by the organism altering that environment to suit its needs (niche construction). In the human case, that reshaping is achieved by technological means derived from the creative activity of the human mind \cite{bronowski,Volti,Har17}.
In the case we are considering here, adaptive selection takes place for specific proteins that have specific functions that  enhance the probability of organisms surviving and reproducing, as is clear from their physiological roles. 

\subsection{A multilevel process}\label{sec;multivelevel}
In biology, this process of adaptation takes place in a contextual way \cite{Cam74,Ell18} through evolutionary emergence  of developmental systems at the lower levels  \cite{Oyaetal01} %(comprised of interacting metabolic networks, signal   transduction networks, and gene regulatory networks)
 that require existence of specific proteins for their function. At the higher levels it leads to development of robust physiological systems \cite{RhoPfl89,AnimalPhysiology} (protected by \href{https://en.wikipedia.org/wiki/Homeostasis}{homeostasis} (\ref{eq:33})) and a plastic brain that can adapt effectively to the physical, ecological, and social environment \cite{bronowski}. An example is that the function $A$ of seeing enabled by the trait $\alpha$ of vision gives great survival advantage to individuals (\cite{Mayr2002}:133,225-228).
Development of vision is a multi-level process, with higher level need $\alpha$ driving lower level selection of structure and function (\cite{AnimalPhysiology}:217-230):
\begin{itemize}
\item The top level need is for a visual system $S$  (\cite{AnimalPhysiology}:252-271) that will enhance survival \cite{Godrfysmith} by reception and interpretation of information from the environment;
\item The next level need is for eyes, an optic tract, thalamus (a \href{http://www.advancedots.com/wp-content/uploads/2016/07/Corticothalamic-feedback-and-sensory-processing.pdf}{relay} station for signals on the way to the cortex), and neocortex to analyse incoming data \cite{AnimalPhysiology};
\item  The next level is a need for photo receptor cells  within the retina (\cite{AnimalPhysiology}:262-271), and neurons and synapses to constitute neural networks to analyse the data; 
\item One then needs specific kinds of proteins to make this all work  \cite{Wag}, for example   \href{https://en.wikipedia.org/wiki/Rhodopsin}{rhodopsins} in Light Harvesting Complexes and \href{https://en.wikipedia.org/wiki/Voltage-gated_ion_channel}{voltage gated ion channel proteins} in the neuronal axons, and in some contexts,  specific proteins for aberration-free lenses \cite{lenses} (see Appendix \ref{sec:visual});
\item So one needs to select for developmental systems \cite{Oyaetal01} to make this all happen:   gene regulatory networks, signal transductions networks, and metabolic regulatory networks, together with the proteins needed to make them work  \cite{Wag}; 
\item   Thus one needs whatever genome will do the job of providing all the above \cite{Wag}. In accordance with the principle of multiple realisation (Section \ref{sec:epigenetic}), there will be numerous genotypes that can do the job. Selection does not take place for specific genes, but for organisms having genotypes that lead to relative advantage \cite{Mayr2002}. 
\end{itemize}

%\subsection{Adaptive selection is a top down process}

%Adeptive processes at macro levels such as learning are  possible because of underlying adaptive selection proceses at micro levels, which select for spacific moecules to exist but are also enabled by existence of such molecules.
%But t
This is all guided by high level needs, as made clear in this example. The environmental niches $\cal{E}$ might be that the animal lives on land, or in the air, or in shallow water, or in deep water. The selection criterion $C$ might be a need to mazimise intensity sensitivity, or edge detection, or  motion detection, or angular resolution, or colour sensitivity; which is most important will depend on the ecological environment. Thus natural selection is a top down process \cite{Cam74,Ell16,Ell18} adapting animals to their environment in suitable ways, thereby altering the details base-pair sequence in DNA. As stated by Stone (\cite{Sto15}:188):
\begin{quote}
``\textit{Evolution is essentially a process in which natural selection acts as a mechanism for transferring information from the environment to the collective genome of the species}''.
\end{quote}
Actually it is doubly a top down process, through the environment $\cal{E}$ creating niches  (opportunities for life) on the one hand, and through the  selection criteria $C$ on the other. Altering either of them alters the micro (genotype) and macro (phenotype) outcomes. The reliability of the resulting systems is because,  as discussed above, biological systems consisting of hierarchically nested, complex networks that 
are extremely robust to extrinsic perturbations are the context within which these adaptive processes take place  \cite{EvnHIerarchy}.
%and underlies origin of information in both the logical and the implementation hierarchies.
%\textbf{Level surfaces}\cite{Wag}

\subsection{What is the role of chance?}
There is a great deal of noise and randomness in biological processes, particularly at the lower levels where molecules live in what has been labelled a `molecular storm' \cite{Hoff12}. The occurrence of this  noise does not mean the outcome is random: rather, it provides an ensemble of variants that is the basis for selection of outcomes according to higher level selection criteria, thus creating order out of disorder in a reliable way \cite{NobleandNoble}, as represented by   (\ref{eq:project1}). Indeed, microbiology thrives on randomness \cite{Hoff12} as does brain function \cite{Gli05,RolDec10}. Statistical randomness between levels provides the material on which selection processes (\ref{eq:select}) can operate, thus allowing higher level needs to direct outcomes. %Quantum uncertainty might also play a role. %(this is unclear until we understand the nature of the quantum measurement problem \cite{Ell2012}). 
Randomness\footnote{With a `hold' mechanism
	that generates functionality and gives the subsequent evolutionary process a direction \cite{NobleandNoble}.} plays a key role in evolution (see \cite{Glymour},\cite{Mayr2002}:252-254,\cite{Kampourakis}:184--191), underlying that vast variety of life on Earth by providing a very  varied set of genotypes on which selection can operate.

\section{Deductive Causation}\label{sec:6}
Deductive causation takes place when effects are the outcome of explicit logical processes, as contrasted to the biological cases discussed so far, where they are processes that are indeed carrying out logical operations but these are implicit in the biology rather than explicit. That is, deductive causation requires mental processes that explicitly consider alternative logical inevitabilities or probabilities and decide outcomes on this basis, for example, ``If I wait till 10am I will miss the bus, so I'd better leave now''. This requires conscious intelligence,\footnote{We note here that these processes can become automated after much practice so that they are intuitive rather than the result of directed mental effort. Nevertheless the nature of the causation is the same.} and certainly occurs in the case of humans. It may also occur to some degree in animals, but we will not enter that debate here: the essential point is that it does indeed occur in the real world, as evidenced by the existence of books, aircraft, digital computers, and all the other products of conscious design \cite{Har17}. It is made possible by the existence of brains (at the macro scale) \cite{Kanetal13} and their  underlying biomolecules such as voltage gated ion channels (at the micro scale) \cite{Scott,Kanetal13}, as discussed in Section \ref{sec:4}.\\

We look in Section \ref{sec:D} at deductive argumentation ${\cal D}$, whose truth is valid independent of contingent facts,  in Section \ref{sec:DE} at evidence based deduction ${\cal D}E$, where the addition of empirical data $E$ leads to conclusions that follow from that evidence via logical deduction ${\cal D}$, and in Section \ref{sec:DEO} at deductively based predictions of outcomes ${\cal D}EO$,  which are used to decide on best choices of actions ${\cal D}EO{\cal C}$ on the basis of logical predictions of outcomes $O$ following from the data $E$ together with choice criteria ${\cal C}$. 

\subsection{Deductive argumentation}\label{sec:D}
Deductive argumentation can be definite or probabilistic. \textit{Definite deductive  arguments} deal with inevitable outcomes of abstract relationships between variables:\footnote{We are not giving a formal definition of logic here, but rather a sketch of how it works. It can be any form of logic that has been discovered by the human mind.}
thus\footnote{Strictly speaking, the word ``necessarily'' is superfluous. We add it for emphasis here and below. A similar remark applies to ``probably'' in (\ref{eq:logic4}).} %formally
%the %included by allowing $T1(\textbf{X})=T2(\textbf{X})$, giving the
 %simpler case
\begin{equation}\label{eq:logic1a}
 %{\texttt{Given\, context\,}\, C,\,
{\cal D}: {\texttt{ IF}\,\, T1(\textbf{X})\,\, \,\texttt{THEN necessarily}\,\,\textbf{} T2(\textbf{Z})},	
\end{equation}
where  $T1(\textbf{X})$ may involve logical operations AND, OR, NOT, and their combinations, or mathematical equalities or inequalities, or both logical and mathematical relations in any combination. %This abroadlysimple elements. 
Thus one might have a conjunction of conditions
\begin{equation}\label{eq:logic1}
%{\texttt{Given\, context\,}\, C,\,
{\cal D}2: {\texttt{ IF}\,\, T1(\textbf{X})\,\, \texttt{\texttt{AND}}\,\, T2(\textbf{Y}),\,\texttt{THEN necessarily}\,\,\textbf{} T3(\textbf{Z})},	
\end{equation}
where $\textbf{X}$, $\textbf{Y}$ and $\textbf{Z}$ may or may not be the same variables. %The conjunction of conditions $T1$ and $T2$ here (which may involve logical operations AND, NOR, NOT, XOR, etc, or mathematical equalities or inequalities) allows building up of complex logic from simple elements. 
These are of the same logical form as (\ref{eq:2}), but the key difference is that in that case, the context was the logic implicitly embodied in biological processes, whereas here the relations refer to explicit logical thought patterns. They may be realised at some moment in a brain, or written down on paper, or recorded in some other way (such as on a black board or a computer screen), %or embodied in computer programs,
 but the patterns themselves are abstract relations with their own internal logic that is independent of whatever specific realisation may occur.\\

 Mathematical examples are the relations 
 \begin{equation}\label{eq:sqrt2}
 \texttt{IF \{X=2\} } \texttt{THEN } \{\sqrt{X} \texttt{ is irrational}\}
 \end{equation}
 which is proved by \href{https://www.homeschoolmath.net/teaching/proof_square_root_2_irrational.php}{algebraic argumentation}, and the partial differential equation result 
\begin{eqnarray}\label{max22}
\texttt{IF \{Eqns.(\ref{eq:maxwell}),(\ref{eq:maxwell1}) hold with } \mathbf{J}  = \rho=0\}, \\ \texttt{ THEN }
\{\texttt{wave solutions } u(x,t) = F(x-ct) 
+ G(x+ct) \texttt{ exist}\} \nonumber
\end{eqnarray}
(which mathematical fact underlies the existence of radios, TV, cellphones, etc).
%, and 
%\begin{equation}\label{galileo}
%\texttt{IF equations } \,\,F=m\frac{dv}{dt}  \texttt{ AND } F = -mg \texttt{ hold, THEN } x = x_0 + v_0 t -\frac{1}{2}gt^2
%\end{equation}
%(which mathematical relation happens to give a good description of a falling body).
 Logical examples are the relations  
\begin{equation}\label{logicalded}
\textit{IF }
\{A \Rightarrow B\} \textit{ AND } \{B\Rightarrow C\}\textit{ THEN }\{A\Rightarrow C\}
\end{equation}
and the combinatorial rules of Boolean logic \cite{boolean}. \\

\textit{Probabilistic logical arguments} deal with likely outcomes on the basis of  statistical evidence, for example:
\begin{equation}\label{eq:logic4}
%{\texttt{Given\, context\,}\, C,\,
{\texttt{ IF }\,\, T1(\textbf{X,P1})\,\, \texttt{AND}\,\, T2(\textbf{Y,P2}),\,\texttt{THEN probably}\,\,\textbf{} T3(\textbf{Z,P3})},	
\end{equation}
where $T1(\textbf{X,P1})$ means $T1$ is valid with probability $P1$, and so on.  A key example is Bayes' Theorem \cite{Sto15}:
\begin{equation}\label{eq:Bayes}
 \texttt{IF } \{P(A)\,\texttt{AND}\,\ P(B|A)\, \texttt{AND}\,\ P(B)\}\, \texttt{ THEN }\, P(A|B) ={\frac {P(B| A)\,P(A)}{P(B)}}\,, % {\displaystyle P(A\mid B)={\frac {P(B\mid A)\,P(A)}{P(B)}}}
\end{equation}
where % $A$ and $B$ are events and $P(B)\neq 0$.Here 
$P(A)$ and $P(B)$ are the probabilities of observing events $A$ and $B$ independent of  each other,
$P(A|B)$  is the conditional probability of observing event $A$ given that $B$ is true, and 
$P(B|A)$ is the conditional probability of observing event $B$ given that $A$ is true. This relation, which is of the form (\ref{eq:logic4}), underlies the learning processes of the predictive brain \cite{Clark,predictive}, enabled by suitable neural structures (\cite{Haw04}; \cite{Tutorial}:\S 2.3-2.5) built from  biomolecules \cite{Scott}. This topic is developed further in Section \ref{sec:bayesian_brain}.

\subsection{The link to data: evidence based deduction}\label{sec:DE}
It may well be that we know that the antecedents in some of these arguments are either  true, or are highly probable, in which case we can  move to evidence based deduction: (\ref{eq:logic1a}) becomes
\begin{equation}\label{eq:deduce1}
{\cal D}E:\texttt{SINCE\,\, T1(\textbf{X})\, \,\texttt{THEN necessarily}\,\,\textbf{} T2(\textbf{Z})},	
\end{equation}
where $T2(Z)$ necessarily follows from $T1(X)$, and we know $T1(X)$ to be true either because we have seen it to be true (there is a dog in  the room), or it is common knowledge (England is near France), or it is an established scientific fact (DNA is a key molecule underlying genetic inheritance), or at least it is a best explanation (established by abduction, i.e. inference to best explanation from observations). For example
\begin{eqnarray}\label{eq:logic442}
%{\texttt{Given\, context\,}\, C,\,
\texttt{ SINCE}\,\, E = m c^2\,\, \,\texttt{THEN binding energy can be made available } \\\texttt{via nuclear fusion of heavy atoms},	\nonumber
\end{eqnarray}
%where the uncertainty relates to technological issues rather than issues of principle, which has no uncertainty (
In other words, because we know special relativity is true, we know we can in principle make nuclear power stations and nuclear bombs. Thus reliable data (the experimental verification of the logically deduced relation $E = mc^2$) relates deductive argumentation to real world possibilities.
Similarly an extension of a simple case of (\ref{eq:logic4}) becomes
\begin{equation}\label{eq:deduce22}
%{\texttt{Given\, context\,}\, C,\,
{\texttt{ SINCE}\,\, T1(\textbf{X,P1})\,\, \,\texttt{THEN probably}\,\,\textbf{} T2(\textbf{Z,P3})},	
\end{equation}
in the probabilistic case, for example
\begin{equation}\label{eq:deduce7}
%{\texttt{Given\, context\,}\, C,\,
{\texttt{ SINCE it is Summer in Oxford}\,\, \, \,\texttt{THEN it will probably rain today.}}	
\end{equation}
The deduction leads to the conclusion that a specific outcome is likely to actually occur. 

\subsection{Deductively based action}\label{sec:DEO}

Following on (\ref{eq:deduce1}) and (\ref{eq:deduce22}), we can deductively determine that specific actions  will inevitably or probably have specific outcomes:	
%If A and B then do C, and it allows further deductions  of the form 
\begin{equation}\label{eq:logic441}
%\texttt{Given\, context\,}\, C,\,
{\cal D}EO: \texttt{ SINCE}\,\, T1 %(\textbf{X,P1})
\,\, \,\texttt{is true THEN %probably
 action A will lead to  outcome O }.%(\textbf{Z,P3})},	
\end{equation}
%where $O$ is the outcome of the action,  
This leads to the basis of deductive choice of best actions:
\begin{equation}\label{eq:logic5}
{\cal D}EO{\cal C}: %\texttt{ %IF}\,\,T1(\textbf{X})\,\, %\texttt{\texttt{AND}}\,\, %T2(\textbf{Y}),
\texttt{ WHEN }\,\, T1 %(\textbf{X,P1})
\,\, \,\texttt{is true\, THEN}\,\,
 \texttt{DO A(\textbf{V}) TO } {\cal C}\texttt{-optimise O} 	
\end{equation}
where ${\cal C}$ is a selection criterion for the best outcome $O_*$,	 and $A(\textbf{V})$ is some action chosen  to alter $O$ via a control variable $V$. The purpose is to produce an optimal outcome $O$ on the basis of a representation of the situation founded on the best available evidence \cite{Papineau}. An example is
\begin{equation}\label{ifthen}
\textit{ WHEN }
\{T > T_0\} \texttt{ THEN } %\{\texttt{set V ON}\}\texttt{ ELSE }
\{\texttt{set V ON\} SO THAT } {\cal C}:\{T< T_0\} %reduce X below } X_0
\end{equation}
which might be part of a computer program implementing feedback control (\ref{eq:33}) to ensure %${\cal C}: 
that temperature $T$ is kept below a critical level $T_0$
 via the cooling control variable $V$.	In the probabilistic case it might be
 \begin{equation}\label{eq:deduce71}
 %{\texttt{Given\, context\,}\, C,\,
 {\texttt{ SINCE \{it is Summer in Oxford\}}\,\, \, \,\texttt{THEN \{take an umbrella\} TO \{keep dry\}.}}	
 \end{equation}
%\begin{equation}\label{eq:logic1a}
 %{\texttt{Given\, context\,}\, C,\,
%{\texttt{ IF}\,\, T1(\textbf{X})\,\, \,\texttt{THEN necessarily}\,\,\textbf{} T2(\textbf{Z})},	
%\end{equation}

When we carry out such deductive argumentation, the abstract logic of the argument ${\cal D}$ (see (\ref{eq:logic1a})) is the causal element determining the nature of the resulting outcomes. The aircraft flies well because we have used explicit deductive mathematical logic ${\cal D}$, together with our knowledge of the laws of fluid dynamics $T1$, to optimize its design O by  running computer aided design packages $A(V)$ representing the aircraft design via variables $V$. We call ${\cal D}$ a `causal element' because of the counter-factual argument \cite{Menzies} that if this abstract logic were different, the outcome would be different. The same applies to ${\cal C}$: if the decision criteria are changed the outcome changes, for example the wing design will be different if the plane is a fighter or an Airbus. This kind of argument is a key part of planning \cite{counterfactual}. \\

In practice (e.g. in economic planning) the argument is often  probabilistic because we can never be absolutely certain of the outcome, due to uncertainty concerning the contextual effects $C$.  
Overall, the import of this section is that
\begin{quote}
\textbf{Deductive causation}: \textit{Logical deductions about scientific, engineering, and social issues can lead to action plans that are causally effective in terms of altering the world. In these cases it is explicit abstract logic ${\cal D}$ realised in brains and/or computers that guides and shapes what happens in highly productive ways \cite{Har17} and hence may be said to be the essential cause of what happens.} 
\end{quote}
This is all possible because of the properties of brains as prediction machines that are also able to make choices between alternatives. The logical operations of deduction ${\cal D}$ and prediction ${\cal D}EO$ take place at the psychological level in the brain  \cite{Ell16}, while being realised at the neural network level through spike chains, at the axon level through ion flows, and at the electronic level through electron movements \cite{Scott}. Each level does work appropriate to the logic at that level, but it is the high level deductive logic ${\cal D}$ that determines what happens in terms of specific outcomes through logically based choices ${\cal D}EOC$  \cite{Ell16}.      

\section{Conclusion: Key Steps in Emergence}\label{sec:7}
How do goal-oriented systems and deductive systems arise out of the  goal-free underlying physics? The main conclusion reached here is that 
\begin{quote}
\textbf{\textit{Biomolecules, and in particular proteins \cite{PetRin09}, form the link between physics and biological causation at the micro scale, underlying emergence of macro-scale purposive entities and deductive causation when incorporated in complex networks. The proteins and the networks must both be selected through processes of adaptive selection. % in order to exist
}}
\end{quote}
We have used ion channels as our main example, because they underlie signal processing in the brain, but there are many other biomolecules that are used in interaction networks to carry out logical operations. In particular \href{https://en.wikipedia.org/wiki/Transcription_factor}{transcription factors}   binding to specific DNA sequences operating by the lock and key molecular recognition mechanism enable logical operations such as AND and NOT (Figure \ref{Fig2}), as do the operation of \href{https://en.wikipedia.org/wiki/Neurotransmission}{synaptic thresholds} associated with \href{https://en.wikipedia.org/wiki/Neuron}{excitatory or inhibitory receptors in neurons} \cite{Kanetal13}. %.,  
%Material basis for information collection and processing
%Sensory systems
%Neural networks to classify, filter, and store
%Neural networks to predict and plan
All of these are adaptively developed, fitting the organism to its environment and based in the lock and key mechanism of supra-molecular biology which enables molecular recognition \cite{Lehn1995,Lehn2007}.

%Thus they are examples of top down realisation in the hierarchy of complexity. 

\subsection{Adaptation and plasticity}
%\textbf{Adaptation and plasticity}
Life depends on adaptation to its environment \cite{CamRee05}. Adaptation takes place at all levels, and at all timescales: evolutionary, developmental, and functional. Brain operations such as learning are context dependent and adaptive \cite{Asbybrain,Ell16}; so are operations of all physiological systems \cite{Nob12}, and all molecular biology processes \cite{GilEpe09}. 
The brain plasticity at macrolevels that underlies our adaptive behaviour \cite{Clark} is enabled at micro levels by biomolecules acting as logical devices (\ref{eq:2}) choosing alternative outcomes on the basis of local and global variables, and altering neuronal connection strengths to enable learning. This is where the key difference from purely physically based interactions occurs. In biology, structure and function go hand in hand \cite{CamRee05}; and this is in particular true in the relation between macro systems and their underlying molecular and cellular structure, which is a two-way process: influences are both bottom-up and top-down \cite{Ell16}; contextual emergence takes place \cite{beimGraben}, and adaptation at macro and miciro levels are intimately related. 

\subsection{Adaptative selection and physics}
The underlying physics enables adaptive selection processes of all kinds to happen, but the physics does not by itself decide what will happen; this is driven by higher level needs (e.g. development of eyes, as discussed above in Section \ref{sec;multivelevel}). Adaptive selection is an emergent biological process \cite{Papineau1}. It acts at all levels: on groups, organisms, systems, cells, interaction networks, genes, and molecules, for each is adapted  to their environment to a greater or lesser extent by adaptive processes on all timescales. 
It is based in physical processes, but is not itself a physical law: it is an essentially biological effect (it is not discussed in physics texts, e.g. \cite{Binney,Blundell}, but is central to biological texts, e.g.   \cite{Mayr2004,CamRee05,Wag11}). In summary,
\begin{quote}
\textbf{\textit{Physics underlies adaptive selection in that it allows the relevant biological mechanisms to work; but the adaptive selection central to biology is not a physical law. It is an emergent biological process that locally confounds the second law  (cast in terms of Shannon entropy), much like Maxwell's daemon does} (see \cite{Friston}).}
\end{quote}
Adaptive selection could not take place without physics, but  unlike physics is a directional process in that it has the outcome of on average increasing fitness. It is irreversible, because species die out in order that others succeed, data is lost when files are deleted and overwritten, and so on; logically this is because it follows from (\ref{eq:project1}) that
\begin{equation}\label{eq:asymmetry}
\{\hat{E}(X)\} \textrm\,\, does\, not\, determine\,\,\{{E}(X)\},
\end{equation} so the data represented by %information 
$\Delta E (X) := \{{E}(X)\}-\{\hat{E}(X)\}
$ is lost in the selection process, unless $\{{E}(X)\}=\{\hat{E}(X)\}$ (but knowing $\{\hat{E}(X)\}$ does not tell you if this was the case or not). This selection process is influenced at higher levels of development by social and psychological influences that crucially shape outcomes \cite{Bir2017}, for example through the development of the language capacity that distinguishes us from the great apes: this symbolic capacity enhances survival because it enables the development of technology \cite{bronowski,Har17}. \\

Adaptive selection is not the same as thermodynamic energy minimisation, although that will play an important part in determining what can happen (and indeed the outcome of some adaptive processes might well be to lead to energy minimisation in cellular or systems processes), nor is it entropy maximisation. It will in general produce broader outcomes related to survival and adaptation to the environment  than energy use optimisation, such as existence of a brain capable of predictive argument and action control. But a brain consumes a large amount of energy: it is an expensive development \cite{Godrfysmith}, it's existence cannot arise simply out of energy minimisation, but rather out of playing a key role in survival prospects. Adaptive selection is a central feature of living systems at all levels of the hierarchy of biological emergence above and including the level of biomolecules,  with outcomes which cannot be deduced by statistical physics methods \cite{Pen79}, nor by a consideration of force laws such as (\ref{eq:maxwell}-\ref{eq:maxforce}), and they are not directly implied by or deducible from the equations of the standard model of particle physics. Darwinian evolution is not implied by physics because physics has no concept of survival of a living being (or for that matter, of a living being), but rather is enabled by physics. Adaptive selection  is not time symmetric (see(\ref{eq:asymmetry})),  despite the fact that the  underlying fundamental physics relevant to biological function is time symmetric (apart from the quantum measurement process \cite{Ell2012}).  

\subsection{The adaptive Bayesian Brain}\label{sec:bayesian_brain}
The deductive processes of Section \ref{sec:6} are determined as valid by the brain through adaptive learning processes leading to logical understanding \cite{Churchland}, enabled by underlying brain plasticity. How does the adaptive selection process of Eqn.(\ref{eq:select}) relate to the %predictive brain understanding \cite{Haw04,Clark}, whereby the brain estimates prediction errors leading to the 
 Bayesian processes of Eqn.(\ref{eq:Bayes} that enable learning)? An outline, based on \cite{Friston}, \cite{SethFreeEnergy},  and particularly \cite{Tutorial}, is as follows:
\begin{itemize}
	\item  Following Clark \cite{Clark} and  Solms and Friston \cite{SolFris18}, the aim 
	of any brain is to minimise the magnitude   $|\textbf{e}| = \left(\sum_i (e_i)^2\right)^{1/2}$ of both short term 
	and long term (vector) prediction error 	$\textbf{e}$ for both interoceptive and proprioceptive domains. The purpose is to produce the best predictions possible on the basis of available data.
	\item The brain uses Bayes' rule (Eqn.(\ref{eq:Bayes})) for the probability $P(\textbf{e})$ to predict %and minimise 
	the likely components of the prediction error $\textbf{e}$
	\item To deal with the influx of data efficiently, the brain deduces most likely values from distributions and deals with them only, throwing away details of the distribution (billions of bits of information are ignored in order to concentrate on what is vital). To determine the most likely value $\phi$ of $v$ which maximizes the posterior distribution $p(v|u)$ on the basis of sensory input $u$, we only have to extremise the log of the numerator in Bayes' theorem (\ref{eq:Bayes}) \cite{Tutorial}. This can happen via  neural circuits shown in Figure \ref{fig:tutorial} (reproduced from Bogacz \cite{Tutorial}). 
	\begin{figure}[h]
		\centering
		\includegraphics[width=0.65\linewidth]{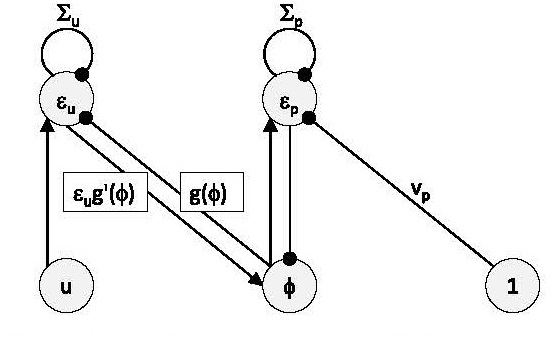}
		\caption{The architecture of the model performing simple perceptual inference. \textit{
				Circles denote neural ``nodes'', arrows denote excitatory connections, while lines
				ended with circles denote inhibitory connections. Labels above the connections
				encode their strength, and lack of label indicates the strength  $1$. Rectangles
				indicate the values that need to be transmitted via the connections they label.} From Bogacz \cite{Tutorial}.}
		\label{fig:tutorial}
	\end{figure}
	\item The effect of these circuits is equivalent to use of the  function \begin{equation}\label{eq:Bayes1}
	F= \ln\, p(\phi) + \ln \,p(u|\phi)
	\end{equation} to determine the most likely value $\phi$ given by Bayes' rule by the gradient ascent method, as per the examples in  \cite{Tutorial}. The key is stated as follows (\cite{Tutorial}:201):
	\begin{quote}
		``\textit{We can find our best guess $\phi$ for} [the inferred value] $v$
			\textit{simply by changing $\phi$ in proportion to the gradient:\footnote{We have conflated his equations (8) and (9).}
			\begin{equation}\label{eq:gradient}
			\dot{\phi} = \frac{\partial F}{\partial \phi} = \frac{v_p-\phi}{\Sigma_p} + \frac{u-g(\phi)}{\Sigma_u}. 
			\end{equation}
			In the above equation $\dot{\phi}$ is the rate of change of $\phi$ with time. Let
			us note that the update of $\phi$ is very intuitive. It is driven by two
			terms: the first moves it towards the mean of the prior,
			the second moves it according to the sensory stimulus, and both
			terms are weighted by the reliabilities of prior and sensory input
			respectively.''}
	\end{quote}
Here $\Sigma_p$, $\Sigma_u$ are the variances in the priors and sensory input respectively, $v_p$ is the prior for the best estimate of the value $\phi$, and $g(\phi)$ is the sensory input $u$ expected to correspond to an object property value $\phi$.   The prediction errors are
	\begin{equation}\label{eq:errors}
\epsilon_p =\frac{v_p-\phi}{\Sigma_p},\,\, 
\epsilon_u=\frac{u-g(\phi)}{\Sigma_u}. 
\end{equation}
		\item Equation (\ref{eq:gradient}) shows how the reliabilities $\Sigma_p$, $\Sigma_u$ are key to the response. In order to evaluate the gradients $\frac{\partial F}{\partial \phi}$, one needs to evaluate the expected precisions, i.e. the variances $\Sigma_p$, $\Sigma_u$. It is plausible \cite{SolFris18}1 that they  correspond to affect, because of the way they change sensitivities and responses \cite{Fri08} as in (\ref{eq:gradient}) - which is what affect does.
\end{itemize}
Overall, what this shows is the following:\footnote{See also \cite{SethFreeEnergy}, last paragraph on page 5.}
\begin{quote}
	 \textit{The Bayesian Brain predictive process \cite{Clark} takes as input a distribution of sensory values with variance $\Sigma_u$ (see \cite{Tutorial}: Figure 1) and, using Baye's Theorem (\ref{eq:Bayes}), selects from them a most likely value $\phi$, rejecting the rest. This projection operation is precisely a selection process of the form Eqn.(\ref{eq:select}).}
\end{quote}
The circuit shown in Figure \ref{fig:tutorial} illustrates how this all can emerge from a network of neurons cnnected by synapses, enabled %at the microlevel
 by the biomolecules discussed in this paper(Section \ref{sec:link}).\\
%\subsection{Contextual nature of biological process}

Why does the brain operate on these Bayesian principles? Because it offers an effective way of updating expectations about the external world \cite{Clark}, and hence reacting in appropriate ways that enhance welfare and survival \cite{Godrfysmith}. The neural networks to implement such abilities (c.f. \cite{Tutorial}) have been developed in the short term via development processes \cite{Wol02,Gil06} that lead to such structures coming into being \cite{Oyaetal01}, and in the long term via the evolutionary adaptive selection processes \cite{CamRee05} that result in genotypes that will lead to such developmental processes coming into existence \cite{Wag11}. Once they exist, they enable deductive argumentation whose outcome is based in the logic itself (see Section \ref{sec:D}) because of their plastic nature and adaptive development \cite{Churchland}. The way energy use supports this functioning is briefly discussed in Appendix \ref{sec:energy}.

\subsection{The major distinctions: three kinds of causation}\label{Sec;3_kinds}
The major difference between physics and life has been characterised above as due to the difference between the immutable impersonal logic of physical causation (\ref{eq:1}) and the  branching purposeful logic of  biological causation (\ref{eq:2}), enabled by biomolecules in general and proteins in particular (Section \ref{sec:link}), %,
 with particularly important kinds of causation  distinguishing life from inanimate matter being  
feedback control (\ref{eq:33}) and adaptation (\ref{eq:select}). \\

% A higher form of causation occurs in intelligent life, when deductively based action (\ref{eq:logic5}) occurs.
 The progression of emergence  is illustrated in Table 1. Inanimate systems are subject only to causation C1. %A key difference between physical and living emergent entities is that a
In all life from cells to organisms to groups to ecosystems, as well as causation C1, causation C2 also occurs, involving logically based branching  (\ref{eq:2}) such  as   homeostasis (\ref{eq:33}) and adaptive selection (\ref{eq:select}).  Thus causation C2  characterises life in general \cite{Hart99} as opposed to inanimate systems. Hence there is a major difference between these two kinds of emergence out of the same basic physical elements \cite{Ell16}. The key thing that enabled causation C2 to emerge historically was the origin of life, when  \href{https://en.wikipedia.org/wiki/Adaptation}{adaptive evolutionary processes} came into being. We \href{https://simple.wikipedia.org/wiki/Origin_of_life}{still do not know how that happened}.  \\
 
 \begin{centering}
 \begin{tabular}{|c|c|c|c|c|}
 \hline 
 \hline 
 & Causation & Agency & Outcome& Reference\\
 
 \hline 
  C1& Physical & Physical laws & Determinist & Eqn.(\ref{eq:1})\\ 
 \hline 
 C2& Biological & Goal-seeking, Selection & Adaptive & Eqns.(\ref{eq:33},\ref{eq:select})\\ 
 \hline 
 C3& Deductive & Logical argument & Planned outcomes & Eqn.(\ref{eq:logic5})\\ 
 \hline 
 \hline
 \end{tabular} \\
 \vspace{0.1in}
 \textbf{Table 1}: The three major forms of causation: physical, biological, and deductive.\footnote{These differences are related to the different kinds of Aristotelian causality, see \cite{Hoff2017a} for a discussion. We will not pursue that issue here.} \textit{Each relyies on the previous one to enable its emergence. }
 \end{centering}
 \vspace{0.1in}
 
 However a  higher form of causation C3 occurs in intelligent life, when deductively based action (\ref{eq:logic5}) occurs, enabling deductive logic \textit{per se} to have causal powers. Emergence of this kind of causation is a  major transition in evolution  \cite{Transitions}. We can characterise intelligent organisms as those that are engaged in causation C3, which enables them to transcend the physical limitations of their bodies through the power of abstract thought, prediction, and planning, that enables technology to develop (so that for example they can fly through the sky, construct huge cities and dams, or make computer systems).   It is this kind of causation (made possible by symbolic systems such as language and mathematics) that underlies the rise of civilisation and the domination of humans over the planet \cite{bronowski,Har17}: we are no longer limited by the strength of our bodies but by the power of our imagination. \\

Note that we are able to say this without having to make any specific comments on the relation between the brain and consciousness. What is indisputable is that deductive causation does indeed take place in the real world, as demonstrated by many examples (such as the existence of aircaft and computers), and is crucially different than the kind of causation characteristic of physics (Section \ref{sec:2}), although it is enabled by that kind of causation (which allows the brain to function as it does \cite{Scott,Kanetal13}).\\

\textbf{Acknowledgement:} We thank Jannie Hofmeyr, Vivienne Russell, Jonathan Birch and Tim Maudlin for helpful comments, and Mark Solms and Karl Friston for useful interactions. We particularly thank Jeremy Butterfield for very useful advice on philosophical aspects of the paper.
 
%\textbf{Brain plasticity} Adaptive processes at macro %levels such as learning are  possible because of underlying adaptive selection proceses at micro levels, which select for spacific moecules to exist but are also enabled by existence of such molecule
%The logical operations at the micro level allow adaptation to occur on evoltionary developmental and functional timescales
%That is the source of biological function and complexity 

%\newpage

%\newpage
%\appendix
\appendix
\section{Technical Notes}\label{Sec:Tech}
\begin{enumerate}
	%\footnote{P M Binder and G F R Ellis (2016) ``Nature, computation and complexity''   \href{http://iopscience.iop.org/article/10.1088/0031-8949/91/6/064004/meta}{\textit{Physica Scripta} \textbf{91}:064004}.}.
	
	\item In parallel to the discussion here, \textit{digital computers} also show a high degree of logical complexity based in the branching logical capacities built into low level devices (\href{https://en.wikipedia.org/wiki/Transistor}{transistors}) that are combined to create \href{https://en.wikipedia.org/wiki/Logic_gate}{logic gates} enabling higher level branching logic and emergence of a tower of virtual machines (see \cite{Ell16}). It is these specific physical micro-structures, in this case products of logical analysis DEOC by the human mind   discussed in Section \ref{sec:6}, that enable the underlying physical laws to generate branching  behaviour embodied in computer programs. Given this physical structure, specific computer programs specify the abstract logic that will be carried out in a particular application, e.g. word processing or image editing. The same physically based microstructure can carry out any logical operations specified \cite{Ell16}. 
	
	\item The \textit{existence} of biomolecules is enabled by  \href{https://en.wikipedia.org/wiki/Covalent_bond}{covalent bonds}, \href{https://en.wikipedia.org/wiki/Hydrogen_bond}{hydrogen bonds}, and \href{https://en.wikipedia.org/wiki/Van_der_Waals_force}{van der Waals forces} \cite{Wat13}. All are based in quantum physics and the electromagnetic interaction (\ref{eq:maxwell}-\ref{eq:maxforce}).

	\item In the case of a complex logical system, you do not get the higher level behaviour by coarse graining, as in the case of determining density and pressure from statistical physics \cite{Pen79}.\footnote{The case of temperature is rather more complex, see \cite{ContextEmerge}}. You get it by \textit{black boxing} and \textit{logical combination}, involving information hiding and abstraction  to characterize the exterior behaviour of a module, see Section \ref{sec:epigenetic}, Ross Ashby's book \textit{An Introduction to Cybernetics} \cite{Asbybrain},  and Giulio Tononi \textit{et al}'s work on Integrated Information Theory \cite{Tononi}. 
	This is particularly clear in the case of digital computer systems, with their explicit apparatus of abstraction, information hiding, and carefully specified module interfaces, see Grady Booch's book \textit{Object Oriented Analysis} \cite{Boo94}.
	
	\item The key physical effect enabling the existence of the biomolecules discussed here, with their functional properties arising out of complex molecular  structures, is the existence of \textit{broken symmetries}. These are what allow quite different kinds of behaviour to emerge at higher levels out of the underlying physical laws, with all their symmetry properties, as explained by Phil Anderson in his famous paper ``More is Different" \cite{Anderson}. Thus the underlying standard model of particle physics is Lorentz invariant, but the emergent biomolecules (such as shown in Figures \ref{fig:voltagegated} and  \ref{fig:potassiumchannel1}) are not. Again the underlying physics relevant to biological functioning is time symmetric, but homeostasis (\ref{eq:33}) and adaptive selection (\ref{eq:select}) (see (\ref{eq:asymmetry})) are not. Hence in the end this is what underlies the difference between physics and biology.
\end{enumerate}

\section{Emergent %and Fundamental
	 Constants and Parameters}\label{sec;appendix}
Fundamental physics is characterised by a set of fixed unchanging constants such as the speed of light $c$ and Planck's constant $\hbar$. By contrast the equations describing emergent systems, whether physical or biological, are characterised by constants and parameters that are contextually dependent. We look at his first as regards physics (Section \ref{sec:Physconst}), then as regards the Hodgkin-Huxley equations (Section \ref{sec: hodg_Hux}), and finally as regards the Bayesian Brain (Section \ref{sec:Bayes_Brain_const}).
 
\subsection{Physics}\label{sec:Physconst}
It is a characteristic of equations describing the basic physics underlying biology that they only contain constants that are invariant across time and space. By contrast, equations for emergent physical effects often have parameters that depend on context. \\

We will not here enter the many deep discussions about the relation between constants in unified theories of fundamental physics, the standard model of particle physics, and the equations of physics that are relevent to the functioning of biology in everyday life \cite{Uzan}. Rather we will just refer to equations (\ref{eq:maxwell})-(\ref{eq:maxforce}) which suffice to cover all non-quantum aspects of the everyday physics underlying biology. Given a choice of physical units, there are two fundamental constants in these equations, namely the speed of light $c$ and the electron charge $e$. If equations only contain these constants,  they represent universal purely physical effects. Additionally  \href{https://en.wikipedia.org/wiki/Boltzmann_constant}{Boltzmann's constant} $k$ is a derivative constant forming a bridge between micro and macro physics, but is also universal in that it is independent of context.
\\

By contrast, Galileo's gravitational constant $g$ in (\ref{eq:maxforce}) is not universal: it is a contextually dependent constant, derived from the mass $M_e$ and radius $R_e$ of the Earth: to a good approximation,\footnote{This leaves out latitude-dependent centrifugal forces} $g = G M_e/R_e^2$, where Newton's gravitational constant $G$ is indeed a universal constant. Nevertheless as far as biology on Earth is concerned, $g$ is a universal constant. 

\subsection{The Hodgkin-Huxley equations}\label{sec: hodg_Hux}
The essential components of Hodgkin-Huxley-type models for the propagation of action potentials due to ionic flows through ion channels  in axon membranes 
%are shown in Figure \ref{fig:hodgkin-huxley} (reproduced from \href{https://en.wikipedia.org/wiki/Hodgkin-Huxley_model}
%{Wikipedia}). \\
can be described by equations for the current $I(t)$ through a cell membrane in terms of controlling parameters $n(t)$, $m(t)$, and $h(t)$.
%\begin{figure}[h]
%	\centering
%	%\includegraphics[width=0.6\linewidth]{350px-Hodgkin-Huxley.png}
%	\caption{Basic components of Hodgkin-Huxley type models. \textit{ These models represent the biophysical characteristic of cell membranes. The lipid bilayer is represented as a capacitance ($C_m$). Voltage-gated and leak ion channels are represented by nonlinear ($g_n$) and linear ($g_L$) conductances, respectively. The electrochemical gradients driving the flow of ions are represented by batteries ($E$), and ion pumps and exchangers are represented by current sources ($I_p$).} Source: \href{https://en.wikipedia.org/wiki/Hodgkin-Huxley_model}
%		{Wikipedia}.}
%	\label{fig:hodgkin-huxley}
%\end{figure}
For a cell with sodium and potassium channels, the total current through the membrane is given by  \cite{HodgHux}:
\begin{eqnarray}\label{eq:H-H1}
I&=&C_{m}{\frac {{\mathrm {d} }V_{m}}{{\mathrm {d} }t}}+{\bar {g}}_{\text{K}}n^{4}(V_{m}-V_{K})+{\bar {g}}_{\text{Na}}m^{3}h(V_{m}-V_{Na})+{\bar {g}}_{l}(V_{m}-V_{l}),\\
{\frac {dn}{dt}}&=&\alpha _{n}(V_{m})(1-n)-\beta _{n}(V_{m})n\\
 {\frac {dm}{dt}}&=&\alpha _{m}(V_{m})(1-m)-\beta _{m}(V_{m})m\\
{\frac {dh}{dt}}&=&\alpha _{h}(V_{m})(1-h)-\beta _{h}(V_{m})h
\end{eqnarray}
where $I$ is the total membrane current per unit area,  $V_m$ is the membrane potential, $C_m$ is the membrane capacitance per unit area,  ${\bar {g}}_{n}$ is the maximal value of the potassium, sodium, and leak conductances per unit area for $n = \kappa$, Na, and $l$ respectively,  related to the conductances by \begin{equation}\label{key}
g_n = {\bar {g}}_{\text{K}}n^{4},\,\, g_{Na}={\bar {g}}_{\text{Na}}m^{3}h,\,\, g_l = {\bar {g}}_{\text{l}}.
\end{equation}
$V_K$ and $V_{Na}$ are the potassium and sodium reversal potentials, and  $V_l$ is the leak reversal potential. Also $\alpha_{i}$ and $\beta _{i}$ are rate constants for the i-th ion channel, which depend on voltage but not time, and $n(t)$, $m(t)$, $h(t)$ are dimensionless functions between 0 and 1 that are associated with potassium channel activation, sodium channel activation, and sodium channel inactivation, respectively. The experimentally determined functions $\alpha_i$  and  $\beta_i$ were given by Hodgkin and Huxley  in their paper \cite{HodgHux} as
\begin{eqnarray}\label{eq:HH-2}
{\displaystyle {\begin{array}{lll}\alpha _{n}(V_{m})={\frac {0.01(10-V_{m})}{\exp {\big (}{\frac {10-V_{m}}{10}}{\big )}-1}}&\alpha _{m}(V_{m})={\frac {0.1(25-V_{m})}{\exp {\big (}{\frac {25-V_{m}}{10}}{\big )}-1}}&\alpha _{h}(V_{m})=0.07\exp {\bigg (}{\frac {-V_{m}}{20}}{\bigg )}\\\beta _{n}(V_{m})=0.125\exp {\bigg (}{\frac {-V_{m}}{80}}{\bigg )}&\beta _{m}(V_{m})=4\exp {\bigg (}{\frac {-V_{m}}{18}}{\bigg )}&\beta _{h}(V_{m})={\frac {1}{\exp {\big (}{\frac {30-V_{m}}{10}}{\big )}+1}}\end{array}}}
\end{eqnarray}

The key point made clearly by Alwyn Scott \cite{Scott} is that none of the equations (\ref{eq:H-H1}) - (\ref{eq:HH-2}), or the constants in those equations, follow either from the standard model of particle physics, or from Maxwell's equations (\ref{eq:maxwell}) together with the Maxwell Force Law  and  Newton's equations (\ref{eq:maxwell1}). However  equations (\ref{eq:H-H1}) - (\ref{eq:HH-2}), with their constants, do indeed determine the ion flows that take place in the squid giant axon, given the biological context of the axon membrane with ion channels. \\

These equations and constants cannot be deduced from physics alone \cite{Scott}. They follow from the membrane structure of axons, whose effects can be modelled by these equations,   where voltage-gated ion channels are represented by nonlinear conductances $g_n$,  and in particular they follow from the size and  shape of voltage-gated ion channels (such as in Figures \ref{fig:voltagegated} and \ref{fig:potassiumchannel1}), which determine the values of the conductances $g_n$. The constants in these equations, describing the properties of the ion channels, underlie squid brain function \cite{Kanetal13}. They are physical quantities arising not from physics per se, but from biological context. Similar constants will underlie brain function in any other species. A detailed discussion of the role of emergence in HH dynamics is given in  \cite{beimGraben}.

\subsection{The Bayesian Brain}\label{sec:Bayes_Brain_const}
At a higher level, the Bayesian brain equations (\ref{eq:Bayes1}), (\ref{eq:gradient}) describe actions at the psychological level, emergent from the underlying action potential propagation in the cortex. They depend on the prediction errors $\Sigma_p(t)$, the sensory variances  $\Sigma_u(t)$, and the predicted sensory input $g(\phi)$ \cite{Tutorial,SethFreeEnergy}. There are no physical constants or parameters in these equations. These quantities all refer to macro brain variables %that do not emerge from the underlying physics in the sense that one cannot deduce them from physcal  : rather they 
that arise out of the learning processes enabled by the brain's plasticity during its interaction with its physical, ecological, and social environment \cite{Clark,Ell16,Godrfysmith}. This is what determines their values. The prediction errors $\Sigma_p(t)$ and prediction functions  $g(\phi)$ are based in psychological understandings, and are  continually updated in reponse to experience \cite{Clark,SethFreeEnergy}. %, rather than physical variables.  
These understandings are thus effective causal variables at the psychological level, affecting future actions. 
%The sensory variances  $\Sigma_u(t)$ are a mixture of biologically determined physical variables affecting perceptions, and affective variables relating to psychological states.   

\section{Neuroenergetics}\label{sec:energy}

\begin{figure}[h]
	\centering
	\includegraphics[width=0.8\linewidth]{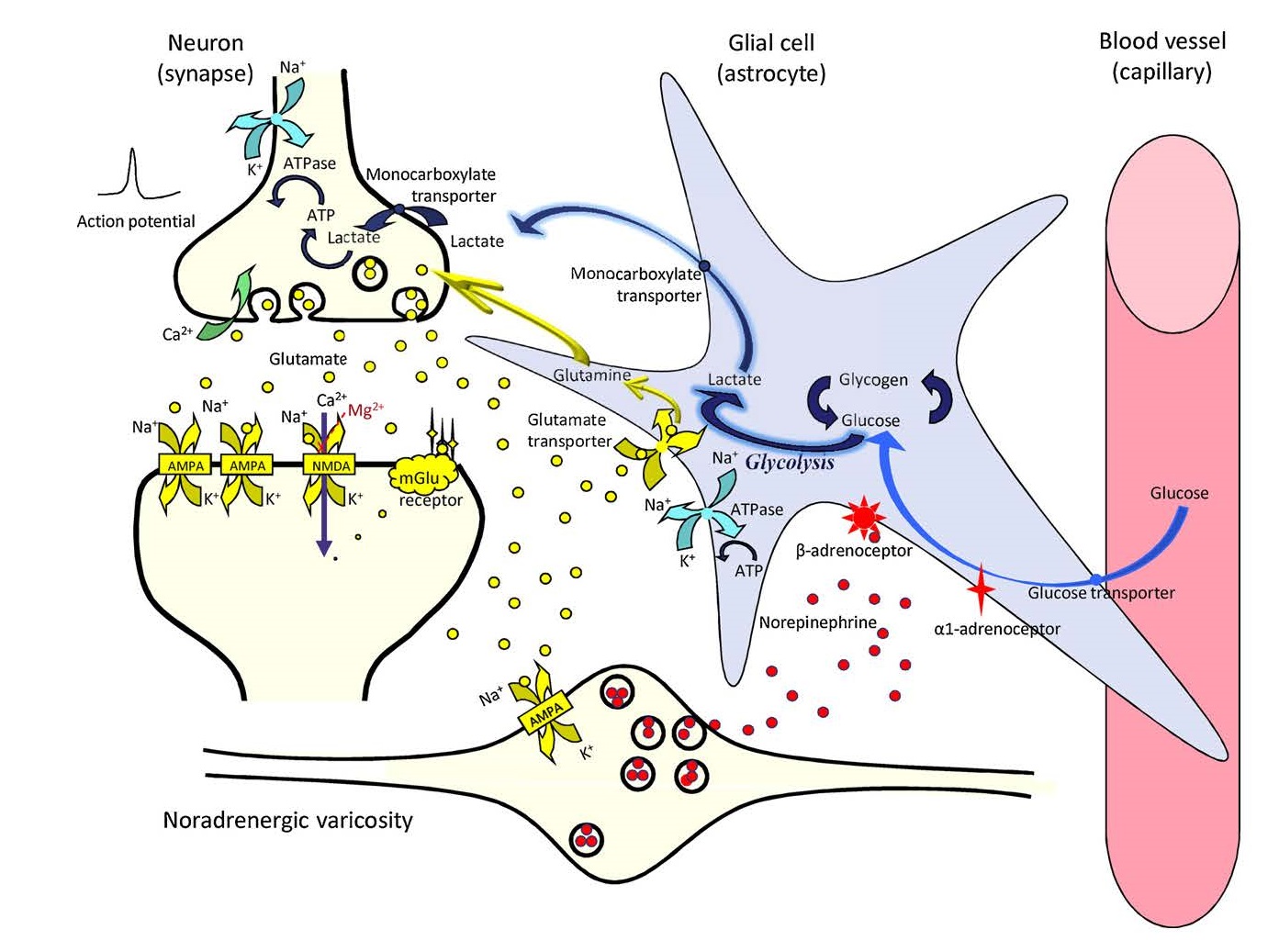}
	\caption{Neuroenergetics \textit{The supply chain for adenosine 
		triphosphate (ATP) production, that powers 
		the neuron.}  These processes are independent of what specific logical operations are being carried out by the brain. From \cite{Russell}.} 
	\label{fig:russellpage05}
\end{figure}
The energy to power brain processes in general, and deductive thought in particular, is provided by metabolic processes whereby glucose in capillaries releases glutamine from astrocytes that in turn supply ATP to neurons \cite{Russell} (see Figure \ref{fig:russellpage05}) 
Release of glutamate 
	(yellow circles) stimulates glucose 
	uptake (blue arrow) and glycolysis in 
	the astrocyte to produce lactate. %These are logically controlled processes as discussed in this paper.
	 The 
	lactate diffuses into the extracellular 
	space, to be absorbed by the neuron for 
	ATP production, %for
	 restoration of 
	ionic gradients, %and resequestration 
	and encapsulation of neurotransmitters. 
	Astrocytes also convert glutamate to 
	glutamine, %which is
	 shuttled to the 
	neurons to restore their pools of 
	neurotransmitters (yellow arrows). 
	Glutamate, acting on AMPA receptors, 
	stimulates norepinephrine release (red 
	circles) from nearby noradrenergic 
	varicosities. These act on $\beta$-adrenoceptors, to further stimulate glucose uptake and glycogenolysis, causing astrocytes to produce more lactate to support sustained neural firing. The rates of these processes are all contextually determined, and will increase in some area when that brain area is more active (which fact can be used for some brain imaging methods). \\
	
	The point of including this outline of \cite{Russell} is that while the neuroenergetics represented by this diagram (which includes numerous feedback control loops implementing homeostasis (\ref{eq:33})) is crucial to brain operation, it functions as it does independent of what specific deductive processing (Section \ref{sec:6}) may be going on in the brain. It enables any Bayesian Brain processes (Section \ref{sec:bayesian_brain}) whatever, irrespective of their cognitive content. Thus these physical processes supplying free energy to the brain underlie the functioning of deductive causation C3 (Section \ref{Sec;3_kinds}), but do not control its nature. They enable it to proceed along its own line of logic ${\cal D}$ (Section \ref{sec:D}). % This is analogous to the way the energy supply to transistors in digital computers does not determine the logic of the computations carried out. This is determined by the algorithms controlling the operations the computer undertakes \cite{Ell16}, for example sorting a list via use of Bubble Sort. Energy or entropy considerations \textit{per se} do not determine these algorithms or outcomes.

\section{Lensing proteins}\label{sec:visual}
\begin{figure}[h]
	\centering
	\includegraphics[width=0.7\linewidth]{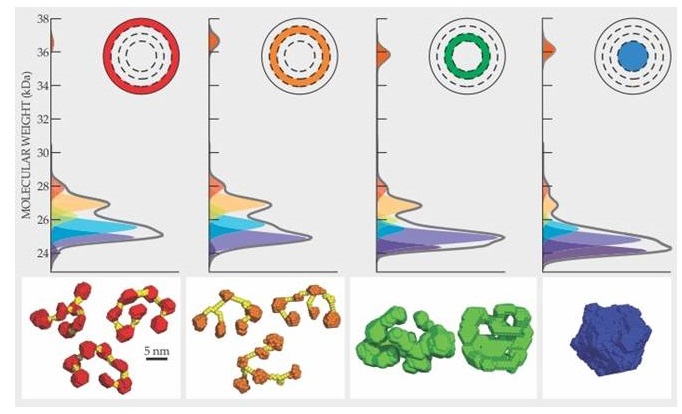}
	\caption{Protein gradients in the squid eye: \textit{Gelatin maintains
		aberration-preventing
		such protein-density gradients. They have been selected for this function.} From \cite{lenses}.}
	\label{fig:lenses}
\end{figure}

Specific proteins have been evolved in order to enable abberation free vision in squids. Developmental processes structure the various proteins in a graded way in the eye \cite{lenses} (see Figure \ref{fig:lenses}, indicating the gradients of the different types of protein).\\

Protein variants combine in radially var.ying ratios in the lens of a squid eye. In this figure, the gray curves show the measured relative molecular-weight distributions of the proteins in the four annular lens layers shown in the inset. The colored peaks are components of a Gaussian fit that takes into account the molecular weights of the variants determined from RNA-sequencing studies. The lower panels show structure models derived from small-angle x-ray scattering data for proteins in each layer. At the periphery of the lens, the proteins form loose chains with each protein linked to two or, occasionally, three others. At the core, the proteins are tightly packed, and each protein is joined to four or more others (\cite{lenses} and references there). \\

The logic underlying this is that if this protein-based structure was not there, the high level need for good visual resolution would not be attained; this is the selection criterion $C$ (Section \ref{sec:selection}). The evolutionary origins of the developmental systems \cite{Oyaetal01} needed to create this structure, and the selection of DNA to code for these specific proteins, is a top-down process from ecological needs \cite{Godrfysmith} to the biomolecular level,  see  Section \ref{sec;multivelevel}. %\\

%Version: 2017/12/21
\end{document}